\documentclass[dvips,nonblindrev]{informs2mod} 

\usepackage{natbib, amsmath, mathrsfs, epsfig, psfrag, subfig, comment, appendix, threeparttable}
\bibpunct[, ]{(}{)}{,}{a}{}{,}%
 \def\bibfont{\small}%
\TheoremsNumberedThrough     
\ECRepeatTheorems

\EquationsNumberedThrough    

\MANUSCRIPTNO{}

\begin{document}


\RUNAUTHOR{Lichtendahl, Grushka-Cockayne, Jose, and Winkler}

\RUNTITLE{Bayesian Ensembles of Binary-Event Forecasts}

\TITLE{Bayesian Ensembles of Binary-Event Forecasts: When Is It Appropriate to Extremize or Anti-Extremize?}

\ARTICLEAUTHORS{
\AUTHOR{Kenneth C. Lichtendahl Jr.}
\AFF{Darden School of Business, University of Virginia, Charlottesville, VA 22903, \EMAIL{lichtendahlc@darden.virginia.edu}}
\AUTHOR{Yael Grushka-Cockayne}
\AFF{Darden School of Business, University of Virginia, Charlottesville, VA 22903, \EMAIL{grushkay@darden.virginia.edu}}
\AUTHOR{{Victor Richmond R. Jose}
\AFF{McDonough School of Business, Georgetown University, Washington, DC 20057, \EMAIL{vrj2@georgetown.edu}}
\AUTHOR{Robert L. Winkler}
\AFF{The Fuqua School of Business, Duke University, Durham, NC 27708, \EMAIL{rwinkler@duke.edu}}
}}

\ABSTRACT{Many organizations face critical decisions that rely on forecasts of binary events. In these situations, organizations often gather forecasts from multiple experts or models and average those forecasts to produce a single aggregate forecast. Because the average forecast is known to be underconfident, methods have been proposed that create an aggregate forecast more extreme than the average forecast. But is it always appropriate to extremize the average forecast? And if not, when is it appropriate to anti-extremize (i.e., to make the aggregate forecast less extreme)? To answer these questions, we introduce a class of optimal aggregators. These aggregators are Bayesian ensembles because they follow from a Bayesian model of the underlying information experts have. Each ensemble is a generalized additive model of experts' probabilities that first transforms the experts' probabilities into their corresponding information states, then linearly combines these information states, and finally transforms the combined information states back into the probability space. Analytically, we find that these optimal aggregators do not always extremize the average forecast, and when they do, they can run counter to existing methods. On two publicly available datasets, we demonstrate that these new ensembles are easily fit to real forecast data and are more accurate than existing methods.}

\HISTORY{\today}
\maketitle

\section{Introduction}
\label{Section: Introduction}
Many organizations face forecasting challenges that involve binary events.  These forecasts are critical to decisions such as the approval of credit (probability of default), the target of a marketing campaign (probability of a click), the recommendation of a drug (probability of having a disease), and the choice of a national security response (probability of a geopolitical event occurring). Often crowds of experts or models issue probabilities for such events. To aggregate the individual probabilities, we offer a new method based on Bayesian principles. This method generalizes several proposed aggregators in the literature, exhibits some structural advantages, and may be more accurate in practice.

Since its introduction by Stone (1961)\nocite{ston:1961}, many researchers have found the linear opinion pool to be an attractive way to aggregate forecasts (DeGroot 1974\nocite{degr:1974}, McConway 1981\nocite{mcco:1981}, Genest and Zidek 1986\nocite{gene:zidek:1986}, DeGroot and Mortera 1991\nocite{degr:mort:1991}). The most popular way to aggregate expert forecasts is to take the average of the experts' forecasts, which is a linear opinion pool with equal weights (Clemen and Winkler 1986\nocite{clem:wink:1986}, Larrick and Soll 2006\nocite{larr:soll:2006}). In aggregating binary-events forecasts, Winkler and Poses (1993, p. 1533)\nocite{wink:pose:1993} state, ``Simple averages of forecasts seem to work as well as or better than fancier combining methods."

Nonetheless, Hora (2004)\nocite{hora:2004} and Ranjan and Gneiting (2010)\nocite{ranj:gnei:2010} show that the linear opinion pool is underconfident.  In the context of binary events, the linear opinion pool is, on average, not extreme enough.  To address this problem, Ranjan and Gneiting (2010)\nocite{ranj:gnei:2010} propose a method that extremizes the linear opinion pool by pushing it closer to its nearer extreme. Many others have employed schemes to extremize the average forecast (Karmarkar 1978\nocite{karm:1978}, Erev et al. 1994\nocite{erev:wall:bude:1994}, Ariely et al. 2000\nocite{arie:au:bend:rand:2000}, Shlomi and Wallsten 2010\nocite{shlo:wall:2010}, Turner et al. 2014\nocite{turn:stey:merk:bude:wall:2014}, Mellers et al. 2014\nocite{mell:unga:baro:2014}, Baron et al. 2014\nocite{baro:mell:tetl:2014}, Satop\"{a}\"{a} et al. 2014\nocite{sato:baro:fost:mell:tetl:unga:2014}). Extremizing methods, such as the logit aggregator, are now used as benchmarks in practice (IARPA Geopolitical Forecasting Challenge 2018\nocite{iarp:2018}). Baron et al. (2014, p. 134)\nocite{baro:mell:tetl:2014} note that ``If every forecaster said 0.6, and they were using different information, then someone who knew all of this would have a right to much higher confidence." This updating process is consistent with Bayesian principles under certain assumptions and is the main justification for extremizing the average forecast.

These existing methods, however, are heuristics that do not follow from Bayesian principles. Consequently, they risk being sub-optimal. The idea that a decision maker can use Bayesian reasoning to aggregate experts' forecasts goes back at least to Winkler (1968)\nocite{wink:1968} and Morris (1974)\nocite{morr:1974}. ``To the expert, the probability assessment is a representation of his state of information; to the decision maker, the probability assessment is information." (Morris 1974\nocite{morr:1974}, p. 1241)  Many other researchers have proposed models along these lines (Lindley et al. 1979\nocite{lind:tver:brow:1979}, French 1980\nocite{fren:1980}, Clemen 1987\nocite{clem:1987}). Dawid et al. (1995)\nocite{dawi:degr:mort:1995} make an important contribution to the literature on aggregating binary-event forecasts. They introduce the first set of aggregation methods based on Bayesian principles and a fundamental condition regarding calibration. The condition says that if the decision maker hears from only one calibrated expert, he adopts that expert's forecast has his own. Recently, Satop\"{a}\"{a} et al. (2016), building on the work of Dawid et al. (1995)\nocite{dawi:degr:mort:1995} and generalizing an example in Ranjan and Gneiting (2010)\nocite{ranj:gnei:2010}, introduce a probit aggregator that is consistent with Bayesian principles. We call such an aggregator a Bayesian ensemble in the spirit of Chipman et al. (2007)\nocite{chip:geor:mccu:2007}.

In this paper, we significantly enlarge the class of Bayesian ensembles. This larger class has three main structural properties, the combination of which offers advantages over previous methods. First, our Bayesian ensembles incorporate the prior-predictive probability into the aggregate forecast. One can think of the prior-predictive probability as the base rate known to the experts and the decision maker alike. For example, when forecasting rain in Phoenix, Arizona, all parties involved in the aggregation may agree on the base rate at which rain occurs daily, say $10\%$. Second, our Bayesian ensembles are generalized linear models of experts' transformed probabilities where the decision maker can choose a link function beyond those that accompany the logit and probit models. With a customized link function, a generalized linear model can often fit data better. Third, because our Bayesian ensembles are generalized linear models of experts' transformed probabilities, they can be easily and quickly estimated on large datasets using statistical computing software.

With this enlarged class of Bayesian ensembles, we provide new insights into why, when, and how to extremize the average forecast. In this class, it often makes sense to extremize the average forecast, but sometimes it does not. In some cases, it can make sense to anti-extremize, or make the aggregate forecast less extreme than the average forecast. Three different types of information play a crucial role in our understanding of extremizing/anti-extremizing: (i) the prior information known to the decision maker and experts, (ii) each expert's private information, and (iii) the shared information known to all the experts but not the decision maker. When the experts rely on private information only, the decision maker naturally wants to form an aggregate forecast that extremizes the average forecast because each individual forecast contains some weight on the prior information. For example, with two experts, the average forecast double counts the prior information and is pulled too far in the direction of the prior-predictive probability.

When the experts rely on the same amount of information but some of that information is shared information, the decision maker does not have as much new information in the reports to justify the same move away from the prior-predictive probability. Hence, shared information tends to reduce the degree of extremizing. In fact, the presence of shared information can cause a Bayesian ensemble to attenuate the average forecast. In practice, shared information is common, since real-world experts often use similar models or have similar training (Kim et al. 2001\nocite{kim:lim:shaw:2001}, Chen et al. 2004\nocite{chen:fine:hube:2004}, Marinovic et al. 2012\nocite{mari:otta:sore:2013}). If models' forecasts are being combined, the models often pick up on similar features from the training set.

In the next two sections, we introduce two types of Bayesian ensembles. The first type, developed in Section~\ref{Section: Conjugate-Pair Bayesian Ensembles}, involves the use of conjugate pairs of distributions. The conjugate pairs help build intuition for how a Bayesian ensemble emerges from the information experts have. The conjugate-pair Bayesian ensembles also help motivate the assumptions behind our second type of Bayesian ensemble, called the generalized probit ensemble. In Section~\ref{Section: Generalized Probit Ensemble}, we provide the form of the generalized probit ensemble and describe how it can be easily fit to data using a generalized linear model with a custom link function. In Section~\ref{Section: Empirical Studies}, we present two pilot studies where we fit the generalized probit ensemble to large sets of real forecast data. In the first study the goal is to predict loan defaults in Fannie Mae's single-family loan performance data. In the second study, the task is to predict defective used cars in Kaggle's {\it Don't Get Kicked!} competition. The forecasts our ensemble and benchmark methods aggregate come from several leading statistical and machine learning algorithms. The results of our pilot studies suggest that the generalized probit ensemble may be more accurate than several leading aggregation methods in other applications.

\section{Conjugate-Pair Bayesian Ensembles}
\label{Section: Conjugate-Pair Bayesian Ensembles}
In this section, we introduce our first type of Bayesian ensemble. Before introducing this ensemble, we provide a formal definition of extremizing---an important definition that will be used throughout the paper.

\subsection{Extremizing}
Our definition of extremizing is inspired by the definitions of sharpness in Winkler and Jose (2008\nocite{wink:jose:2008}) and Ranjan and Gneiting (2010). Sharper, or more extreme forecasts, are those farther away from their marginal event frequencies. Winkler and Jose (2008) state, ``If climatology $c$ is used as the baseline probability for probability of precipitation forecasts, sharpness should be viewed in terms of shifts from $c$ toward zero or one instead of shifts from 0.5 toward zero or one." In forecasting rain in Phoenix, Arizona where the historical daily frequency of rain is about $10\%$, a forecast of $40\%$ would naturally be considered more extreme than a forecast of $30\%$.

For the following definition, we assume the average forecast is not equal to either the prior-predictive probability (i.e., the forecast based on the prior information only) or the aggregate forecast.

\begin{definition}[Extremizing]
\begin{it}
The aggregate forecast extremizes the average forecast if it is farther away from the prior-predictive probability in the same direction as the average forecast. Otherwise, the aggregate forecast anti-extremizes the average forecast.
\end{it}
\label{Definition 1: Extremizing}
\end{definition}

This definition differs from other definitions in the literature. For example, Baron et al. 2014\nocite{baro:mell:tetl:2014}, Satop\"{a}\"{a} et al. (2014\nocite{sato:baro:fost:mell:tetl:unga:2014}), and Satop\"{a}\"{a} et al. (2016\nocite{sato:pema:unga:2016}) say the aggregate forecast extremizes the average forecast if it is farther away from one-half in the same direction as the average forecast. Under this definition, a forecast of $40\%$ for rain in Phoenix would be considered less extreme than a forecast of $30\%$.

\subsection{Information Structure}
Below we introduce an information structure that describes the information the decision maker and experts have in order to make their forecasts. In this setting, each of $k\geq 2$ experts issues a probability forecast for the binary event $y$ given her private sample information $\boldsymbol{x}_i$ and the shared sample information $\boldsymbol{x}_{s}$. The event $y$ equals 1 if the event occurs, or it equals 0 if the event does not occur. We denote expert $i$'s forecast $P(y=1|\boldsymbol{x}_i,\boldsymbol{x}_{s})$ by $p_i$.  After hearing from his $k \geq 2$ experts, the decision maker aggregates the experts' forecasts into a single forecast.

A decision maker's aggregate forecast is a Bayesian ensemble, denoted by $\hat{p}$, if it is the posterior-predictive probability $P(y=1|p_1,\ldots,p_k)$ derived from the joint distribution of $(p_1,\ldots,p_k,y)$ using Bayes' Theorem. This aggregate forecast is optimal because using any other forecast would yield a worse score when evaluated by a proper scoring rule (Gneiting and Raftery 2007). Later in this section we compare several Bayesian ensembles to the average forecast, $\bar{p} = (1/k)\sum_{i=1}^k p_i$, to see when these ensembles extremize (or anti-extremize) the average forecast.

Suppose there is an exchangeable sequence of data points:
\begin{equation}
\begin{split}
(\underbrace{x_1,\ldots,x_{n_1}}_{\stackrel{\mbox{\footnotesize Expert 1's}}{\stackrel{\mbox{\small private sample}}{\vspace{1cm}\mbox{\small of size $n_1$}}}},
\underbrace{x_{n_1 + 1},\ldots,x_{n_1 + n_2}}_{\stackrel{\mbox{\footnotesize Expert 2's}}{\stackrel{\mbox{\small private sample}}{\vspace{1cm}\mbox{\small of size $n_2$}}}},
\ldots,
\underbrace{x_{N_{k-1}+1},\ldots, x_{N_k}}_{\stackrel{\mbox{\footnotesize Expert k's}}{\stackrel{\mbox{\small private sample}}{\vspace{1cm}\mbox{\small of size $n_k$}}}},
\underbrace{x_{N_{k}+1},\ldots, x_{N_{k}+n_{s}}}_{\stackrel{\mbox{\footnotesize Shared}}{\stackrel{\mbox{\small sample}}{\vspace{1cm}\mbox{\small of size $n_{s}$}}}},
\underbrace{x_{N_{k}+n_{s}+1}}_{\stackrel{\mbox{\footnotesize Related to}}{\stackrel{\mbox{\small the binary}}{\vspace{1cm}\mbox{\small event $y$}}}}).
\label{Equation: Expert data with Shared Info}
\end{split}
\end{equation}
Expert $i$ sees the private sample $\boldsymbol{x}_i = (x_{N_{i-1} + 1},\ldots,x_{N_i})$ of size $n_i$ for $i=1,\ldots,k$, where $N_i = \sum_{l=1}^i n_l$. The shared information---known to all the experts, but not the decision maker---is the sample $\boldsymbol{x}_{s} = (x_{N_{k} + 1},\ldots,x_{N_{k}+n_{s}})$ of size $n_{s}$.

The final data point $x_{N_{k}+n_{s} +1}$, abbreviated as $x$, is related to the binary event $y$, which is what the decision maker ultimately cares about. If $x$ is in the event occurrence set $A$, then $y$ equals 1; otherwise, $y$ equals 0. For example, if $x$ is a Bernoulli random variable, the set $A$ might simply be $\{1\}$. Alternatively, if $x$ is a normal random variable, the set $A$ might be the interval $(0,\infty)$.

Data points in the sequence are independent and identically distributed according to a likelihood from a regular, one-parameter exponential family with probability mass or density function $f(x_j|\theta) = a(x_j)b(\theta)\exp(c(\theta)h(x_j))$. The parameter $\theta$ is distributed according to a conjugate prior $f(\theta) = [K(\tau_0,\tau_1)]^{-1}[b(\theta)]^{\tau_0}\exp(c(\theta)\tau_1)$, where $\tau_0$ and $\tau_1$ are the prior's hyperparameters and $K(\tau_0,\tau_1)$ is its normalizing constant. The joint distribution of $(\theta, \boldsymbol{x}_1,\ldots,\boldsymbol{x}_{k},\boldsymbol{x}_{s},x)$ and the event occurrence set $A$ are common knowledge among the decision maker and the experts. The prior/likelihood pair of distributions that describes this joint distribution is called a conjugate pair (Raiffa and Schlaifer 1961\nocite{raif:schl:1961}, Bernardo and Smith 2000). Hence we call the Bayesian ensemble that a conjugate pair generates a conjugate-pair Bayesian ensemble.

Based on these assumptions, the following function generates a set of predictive distributions---one for the prior-predictive, one for each expert's posterior-predictive, and one for the decision maker's posterior-predictive. We call this function the predictive generating function:
\begin{equation}
\begin{split}
F_{n}(t) = \int_{x \in A} a(x)\frac{K(\tau_0 + n + 1, t + h(x))}{K(\tau_0 + n, t)} \, dx,
\label{Equation: Predicitive Generating Function}
\end{split}
\end{equation}
where $n$ is the relevant sample size and $t$ is the relevant sufficient statistic (Bernardo and Smith 2000\nocite{bern:smit:2000}). Note that when the random variable $x$ is discrete, the integral in \eqref{Equation: Predicitive Generating Function}, and other integrals like it throughout the paper, naturally become sums. With $n=0$ and $t=\tau_1$, $F_{0}(\tau_1)$ is everyone's prior-predictive probability $P(y=1)$, denoted by $p_0$. With $n=n_i+n_{s}$ and $t=t_i+t_{s}$ where $t_i = \sum_{j = N_{i-1} + 1}^{N_i} h(x_j)$ and $t_{s} = \sum_{j = N_{k} + 1}^{N_{k}+n_s} h(x_j)$, $F_{n_i+n_{s}}(\tau_1 + t_i + t_{s})$ is expert $i$'s posterior-predictive probability $P(y=1|\boldsymbol{x}_i,\boldsymbol{x}_{s})$.

With $n=N_{k}+n_{s}$ and $t=\sum_{i=1}^{k} t_i + t_s$,  $F_{N_{k}+n_{s}}(\tau_1 + \sum_{i=1}^{k} t_i + t_s)$ is the decision maker's posterior-predictive probability $P(y=1|\boldsymbol{x}_1,\ldots, \boldsymbol{x}_{k}, \boldsymbol{x}_{s})$, as if he had access to all the experts' private and shared information. In reality, the decision maker only hears $p_i$ from expert $i$, so the best he can do is infer $(t_1,\ldots,t_{k}, t_{s})$, the sufficient statistics for $(\boldsymbol{x}_1,\ldots,\boldsymbol{x}_{k},\boldsymbol{x}_{s})$, from $(t_1+t_{s},\ldots,t_{k}+t_{s})$. He can learn each $t_i+t_{s}$ from $p_i$ if $F_{n}$ is invertible. If it is invertible, then $t_1+t_{s} = F_{n_i+n_{s}}^{-1}(p_i) - \tau_1$.

\subsection{Conjugate Pairs with Private Information Only}
In the case of private information only ($n_{s}=0$), the conjugate-pair Bayesian ensemble is a generalized additive model in the experts' probabilities and a generalized linear model in the experts' sufficient statistics. For this result, we need the following two definitions. A generalized additive model links the conditional expectation of a quantity of interest $y$ to an additive function of some covariates $(q_1,\ldots,q_k)$: $E[y|q_1,\ldots,q_k] = g^{-1}(g_0 + g_1(q_1) + \cdots + g_k(q_k))$ where $g$ is the link function, $g_0$ is a constant, and each $g_i$ for $i = 1,\ldots,k$ is a smooth function (Hastie and Tibshirani 1986\nocite{hast:tibs:1986}). A generalized linear model is a general additive model where each $g_i$ is a linear function (Nelder and Wedderburn 1972\nocite{neld:wedd:1972}). Proofs of this and other results appear in the Appendix.

\begin{proposition}[Private Information Only]
\begin{it}
Assume only private information is available to the experts ($n_i > 0$ for $i=1,\ldots,k$ and $n_{s}=0$) and the predictive generating function $F_{n}(t)$ in \eqref{Equation: Predicitive Generating Function} is strictly monotonic in $t$. Then the conjugate-pair Bayesian ensemble of the experts' probabilities is a generalized additive model:
\begin{equation}
\begin{split}
\hat{p} = P(y=1|p_1,\ldots,p_k) = F_{N_k}\bigg(-(k-1)F_{0}^{-1}(p_0)  + \sum_{i =1}^k F_{n_i}^{-1}(p_i)\bigg),
\label{Equation: CP Bayesian Ensemble of Probabilities}
\end{split}
\end{equation}
where $p_0 = F_{0}(\tau_1)$, $p_i = F_{n_i}(\tau_1 + t_i)$, and $t_i = \sum_{j = N_{i-1} + 1}^{N_i} h(x_j)$. Also, the Bayesian ensemble of the experts' sufficient statistics is a generalized linear model:
\begin{equation}
\begin{split}
P(y=1|t_1,\ldots,t_k) = F_{N_k}\bigg(\tau_1  + \sum_{i =1}^k t_i\bigg).
\label{Equation: CP Bayesian Ensemble of Information States}
\end{split}
\end{equation}
\label{Proposition: Conjugate-Pair Ensemble with Private Information Only}
\end{it}
\end{proposition}

This result provides a large class of Bayesian ensembles. The class is as large as the class of regular, one-parameter exponential families. Any ensemble in this class is a generalized additive model of experts' probabilities that first transforms the experts' probabilities into their corresponding information states, then linearly combines these information states, and finally transforms the combined information states back into the probability space.

Below we provide four examples of conjugate-pair Bayesian ensembles. Our first example appears in Dawid et al. (1995)\nocite{dawi:degr:mort:1995}, and our third example is a variant of the models studied in Ranjan and Gneiting (2010)\nocite{ranj:gnei:2010} and Satop\"{a}\"{a} et al. (2016)\nocite{sato:pema:unga:2016}. The other two examples are new. For details on the derivation of each example's ensemble, e.g., each example's $\tau_1$, $h(x_j)$, and $F_n(t)$, see the Appendix.

\begin{example}[Beta/Bernoulli Pair]
\begin{rm}
Let $x_j$ given $\theta$ be drawn from a Bernoulli distribution with probability $\theta$. The conjugate prior for this likelihood is the beta distribution with shape parameters $\alpha$ and $\beta$. Suppose $A = \{1\}$ corresponds to the event that a future borrower defaults on a loan.  With the event occurrence set $A = \{1\}$, this conjugate pair leads to the Bayesian ensemble
\begin{equation}
\begin{split}
\hat{p} = \bigg(1-\sum_{i=1}^k w_{n_i} \bigg) p_0+ \sum_{i=1}^k w_{n_i} p_i,
\nonumber
\end{split}
\end{equation}
where $w_{n_i} = (\alpha + \beta + n_i)/(\alpha + \beta + N_k)$.\Halmos
\label{Example: Beta/Bernoulli Pair}
\end{rm}
\end{example}

\begin{example}[Gamma/Poisson Pair]
\begin{rm}
Let $x_j$ given $\theta$ be drawn from a Poisson distribution with rate $\theta$. The conjugate prior for this likelihood is the gamma distribution with shape $\alpha$ and rate $\beta$, denoted by $\theta \sim \mathit{Ga}(\alpha, \beta)$. Suppose $A = \{0\}$ corresponds to the event that a piece of equipment, with exponentially distributed interarrival times of breakdowns, does not break down in the next year.  With $A = \{0\}$, this conjugate pair leads to the Bayesian ensemble
\begin{equation}
\begin{split}
\hat{p} = \exp\bigg(-(k-1)\frac{v_{N_k}}{v_{0}}\log(p_0) + \sum_{i=1}^k \frac{v_{N_k}}{v_{n_i}}\log(p_i)\bigg),
\nonumber
\end{split}
\end{equation}
where $v_{n} = \log((\beta + n)/(\beta + n + 1))$.\Halmos
\label{Example: Gamma/Poisson Pair}
\end{rm}
\end{example}

\begin{example}[Normal/Normal Pair]
\begin{rm}
Let $x_j$ given $\theta$ be drawn from a normal distribution with mean $\theta$ and variance $\sigma^2$. The conjugate prior for this likelihood is another normal distribution with mean $\theta_0$ and variance $\sigma_0^2$. Suppose $A = (0, \infty)$ corresponds to the event that a new product makes a profit in its first year.   With $A = (0, \infty)$, this conjugate pair leads to the Bayesian ensemble
\begin{equation}
\begin{split}
\hat{p} = \Phi\bigg(-(k-1)\sqrt{\frac{v_{0}}{v_{N_k}}}\Phi^{-1}(p_0) + \sum_{i=1}^k \sqrt{\frac{v_{n_i}}{v_{N_k}}} \Phi^{-1}(p_i) \bigg),
\nonumber
\end{split}
\end{equation}
where $\Phi$ is the cumulative distribution function (cdf) of the standard normal distribution and $v_n = (\sigma^2/\sigma_0^2+n)(\sigma^2/\sigma_0^2+n +1)\sigma^2$. We call this ensemble a probit ensemble because the inverse link function in this generalized linear model is the standard normal cdf.\Halmos
\label{Example: Normal/Normal Pair}
\end{rm}
\end{example}

\begin{example}[Generalized-Gamma/Gumbel Pair]
\begin{rm}
Let $x_j$ given $\theta$ be drawn from a Gumbel distribution with location $\theta$ and scale $\sigma$. The conjugate prior for this likelihood is the reflection of the generalized gamma distribution in Ahuja and Nash (1967, Equation 2.7)\nocite{ahuj:nash:1967}: $\exp(\theta/\sigma) \sim \mathit{Ga}(\alpha, \beta)$. Suppose $A = (-\infty,0)$ corresponds to the event that a hedge-fund manager's best investment makes a loss in some year. With $A = (-\infty,0)$, this conjugate pair leads to the Bayesian ensemble
\begin{equation}
\begin{split}
\hat{p} = \bigg(\frac{-(k-1)p_0^{v_{0}}/(1-p_0^{v_{0}}) + \sum_{i=1}^k p_i^{v_{n_i}}/(1-p_i^{v_{n_i}})}{1-(k-1)p_0^{v_{0}}/(1-p_0^{v_{0}}) + \sum_{i=1}^k p_i^{v_{n_i}}/(1-p_i^{v_{n_i}})}\bigg)^{1/v_{N_k}},
\nonumber
\end{split}
\end{equation}
where $v_{n} = 1/(\alpha + n)$.\Halmos
\label{Example: Generalized-Gamma/Gumbel Pair}
\end{rm}
\end{example}

The ensembles in Examples~\ref{Example: Beta/Bernoulli Pair} and \ref{Example: Normal/Normal Pair} are depicted in Figure~1a and 1c, respectively. In both figures, we hold $p_1$ fixed and vary $p_2$. For Figure~1a, we assume $\alpha=\beta=1$, two experts each privately see two data points, and expert~1 reports $p_1 = 3/4$. For Figure~1c, we assume $\sigma =\sigma_0=1$, $\theta_0=-1.25$, two experts each privately see two data points, and expert~1 reports $p_1 = F_{n_1}(\tau_1) = F_2(-1.25)\approx 0.36$.

\begin{figure}[h!]
\begin{center}
\vspace{0cm}
\subfloat[Beta/Bernoulli (Private Only).]{\includegraphics[width=7cm]{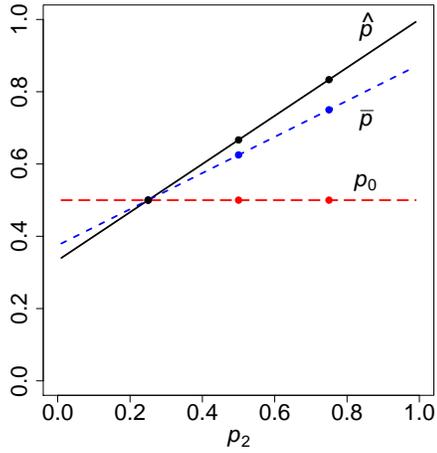}}
\subfloat[Beta/Bernoulli (Private and Shared).]{\includegraphics[width=7cm]{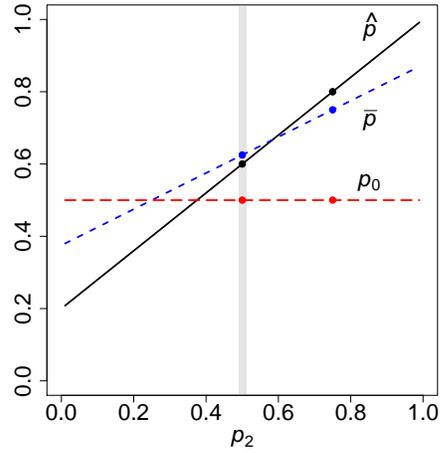}}
\\
\subfloat[Normal/Normal (Private Only).]{\includegraphics[width=7cm]{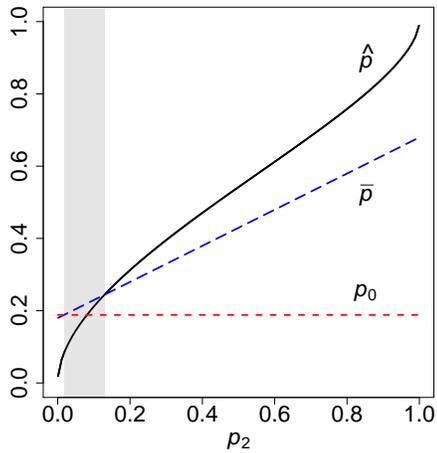}}
\subfloat[Normal/Normal (Private and Shared).]{\includegraphics[width=7cm]{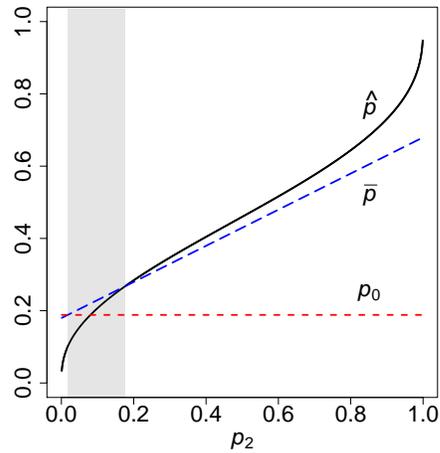}}
\vspace{1cm}
\caption{Conjugate-Pair Bayesian Ensembles from Examples~\ref{Example: Beta/Bernoulli Pair} and \ref{Example: Beta/Bernoulli Pair with Private and Shared Information} (Beta/Bernoulli) and Examples~\ref{Example: Normal/Normal Pair} and \ref{Example: Normal/Normal Pair with Private and Shared Information} (Normal/Normal).}
\label{Figure: Examples of Conjugate-Pair Bayesian Ensembles}
\end{center}
\end{figure}

With exchangeable experts (i.e., each $n_{i}=n_1$), the following strict inequality holds for the beta/Bernoulli ensemble in Example~\ref{Example: Beta/Bernoulli Pair}:
\begin{equation}
\begin{split}
\hat{p} &= (1-k w_{n_1}) p_0+ \sum_{i=1}^k w_{n_1} p_i = (1-k w_{n_1}) p_0+  kw_{n_1} \bar{p} < (1-k w_{n_1}) \bar{p} +  kw_{n_1} \bar{p} = \bar{p},
\label{Equation: Beta/Bernoulli Ensemble Always Extremizes}
\end{split}
\end{equation}
for $\bar{p} < p_{0}$. Similarly, for $\bar{p} > p_{0}$, the inequality in \eqref{Equation: Beta/Bernoulli Ensemble Always Extremizes} is reversed. Thus, with private information only, the beta/Bernoulli ensemble always extremizes the average forecast. The probit ensemble, however, sometimes does not. Its region of anti-exremizing (in gray) is sizeable.

The intuition for why a Bayesian ensemble tends to extremize the average forecast is that the experts' forecasts each have some weight on the prior information, but when their forecasts are combined, the aggregate forecast only needs that weight once. This intuition explains the coefficient $-(k-1)$ in front of the term $F_{0}^{-1}(p_0)$ in \eqref{Equation: CP Bayesian Ensemble of Probabilities}. In other words, extremizing tends to be the result of removing the redundant weight on the prior information. Because the experts' probabilities are first non-linearly transformed inside the probit ensemble, it does not always extremize the average forecast. In the next subsection, we compare these two ensembles to their shared information counterparts to see what effect this information has on extremizing/anti-extremizing.

\subsection{Conjugate Pairs with Private and Shared Information}
The following result generalizes Proposition~\ref{Proposition: Conjugate-Pair Ensemble with Private Information Only} to include the presence of shared information.

\begin{proposition}[Private and Shared Information]
\begin{it}
Assume private and shared information are available to the experts ($n_i > 0$ for $i=1,\ldots,k$ and $n_{s}>0$) and the predictive generating function $F_{n}(t)$ in \eqref{Equation: Predicitive Generating Function} is strictly monotonic in $t$. Then the conjugate-pair Bayesian ensemble of the experts' probabilities is given by
\begin{equation}
\begin{split}
\hat{p} &= \int F_{N_{k}+n_s}\bigg(-(k-1)F_{0}^{-1}(p_0)  + \sum_{i =1}^k F_{n_i+n_{s}}^{-1}(p_i) -(k-1)t_{s}\bigg) f(t_{s}|p_1,\ldots,p_k) \ dt_s,
\label{Equation: CP Bayesian Ensemble of Probabilities with Shared Info}
\end{split}
\end{equation}
where $p_0 = F_0(\tau_1)$, $p_i = F_{n_i+n_{s}}(\tau_1 + t_i + t_{s})$, and $f(t_{s}|p_1,\ldots,p_k)$ is the probability mass or density function of the shared sample's sufficient statistic $t_{s}$ conditional on the experts' reported probabilities.
\label{Proposition: Conjugate-Pair Ensemble with Private and Shared Information}
\end{it}
\end{proposition}

In the presence of shared information, the decision maker can, at most, deduce each expert's sufficient statistic $t_1 + t_{s}$ from $p_1$: $t_1+t_{s} = F_{n_i+n_{s}}^{-1}(p_i) - \tau_1$. So, in the integral (or sum) in \eqref{Equation: CP Bayesian Ensemble of Probabilities with Shared Info}, $f(t_{s}|p_1,\ldots,p_k)$ is equal to $f(t_{s}|t_1 + t_{s},\ldots,t_k + t_{s})$, which is not always tractable. The example below illustrates a situation where this conditional distribution is quite simple to evaluate.

\begin{example}[Beta/Bernoulli Pair with Private and Shared Information]
\begin{rm}
Assume $\alpha=\beta=1$, $x_1$ and $x_2$ are private information seen by experts 1 and 2, respectively, $x_3$ is shared information, and $x_4$ is related to the binary event $y$. The binary event $y$ (default or not) is 1 if $x_4=1$ and is 0 otherwise. The prior-predictive probability $p_0=1/2$. Suppose expert~1 reports $p_1 = 3/4$. Given this report, the decision maker can deduce that $x_1 + x_3 =2$ (or $x_1=1$ and $x_3=1$): he knows expert~1 saw two defaults in her private and shared information. Based on expert~1's report, the decision maker also knows that $x_2 + x_3$ can only be either 1 or 2, leading to either $p_2 = 1/4$ or $p_2 = 1/2$. In either case, $f(t_{s}|p_1,\ldots,p_k)$ is a point mass because $P(x_3 = 1|x_1 + x_3 =2, x_2 + x_3 = 1) = P(x_3 = 1|x_1 + x_3 =2, x_2 + x_3 = 2) = 1$. Consequently, in this example, the beta/Bernoulli Bayesian ensemble with private and shared information becomes
\begin{equation}
\begin{split}
\hat{p} &= P(y=1|p_1=3/4,p_2) = \frac{4p_2 + 1}{5}.\Halmos
\label{Equation: Beta/Bernoulli Bayesian Ensemble with Shared Info and One Term}
\end{split}
\end{equation}
\label{Example: Beta/Bernoulli Pair with Private and Shared Information}
\end{rm}
\end{example}

In Figure~\ref{Figure: Examples of Conjugate-Pair Bayesian Ensembles}, we compare Example~\ref{Example: Beta/Bernoulli Pair with Private and Shared Information}'s ensemble with private and shared information to Example~\ref{Example: Beta/Bernoulli Pair}'s ensemble with private information only. In both examples, the experts see two data points the decision maker does not see. Turning one piece of private information into shared information, as we move from Figure~1a to Figure~1b, we see a reduction in the degree of extremizing. In Figure~1b, we also see a point of anti-extremizing (in gray) when $p_2 = 0.5$.

The conditional distribution $f(t_{s}|p_1,\ldots,p_k)$ is not always the point mass we see in Example~\ref{Example: Beta/Bernoulli Pair with Private and Shared Information}. If $p_1$ were 1/2 there, then the decision maker could be uncertain about the event $x_3=1$ and the Bayesian ensemble would be a mixture. In this case, the sum in \eqref{Equation: CP Bayesian Ensemble of Probabilities with Shared Info} would contain two terms when expert~2 reports $1/2$, rather than the single term in \eqref{Equation: Beta/Bernoulli Bayesian Ensemble with Shared Info and One Term}. In general, this integral (or sum) is difficult to evaluate. For the normal-normal pair though, the Bayesian ensemble is not a mixture and has a tractable form.

\begin{proposition}[Private and Shared Information: Normal/Normal Pair]
\begin{it}
Assume private and shared information are available to the experts ($n_i > 0$ for $i=1,\ldots,k$ and $n_{s}>0$) and the conditions of Example~\ref{Example: Normal/Normal Pair}'s normal/normal conjugate pair hold. Then the normal/normal conjugate-pair leads to the Bayesian ensemble of the experts' probabilities:
\begin{equation}
\begin{split}
\hat{p} &= \Phi\bigg( \frac{-(k-1)\sqrt{v_{0}}\Phi^{-1}(p_0) + \sum_{i =1}^k \sqrt{v_{n_i+n_{s}}}\Phi^{-1}(p_i) - (k-1) E[t_{s}|p_1,\ldots,p_k]}{ \sqrt{(k-1)^2 \mathit{Var}[t_{s}| p_1,\ldots,p_k] + v_{N_{k}+n_s}}} \bigg),
\label{Equation: Probit Ensemble of Probabilities with Shared Info}
\end{split}
\end{equation}
where $v_n = (\sigma^2/\sigma_0^2+n)(\sigma^2/\sigma_0^2+n +1)\sigma^2$. The conditional mean $E[t_{s}|p_1,\ldots,p_k]$ is linear in $(\Phi^{-1}(p_1),\ldots,\Phi^{-1}(p_k))$, and the conditional variance $\mathit{Var}[t_{s}| p_1,\ldots,p_k]$ is constant in $(p_1,\ldots,p_k)$. (The conditional moments are given in the proof of this result.)
\label{Proposition: Normal/Normal Pair Ensemble with Private and Shared Information}
\end{it}
\end{proposition}

Similar to the probit ensemble with private information only, the ensemble in the result above is a linear combination of $(\Phi^{-1}(p_0),\Phi^{-1}(p_1),\ldots,\Phi^{-1}(p_k))$ inside the standard normal cdf. Therefore, we call the ensemble in \eqref{Equation: Probit Ensemble of Probabilities with Shared Info} a probit ensemble as well. The two probit ensembles differ only in the weights in their respective linear combinations. These weights determine the degree of extremizing, as the next example illustrates.

\begin{example}[Normal/Normal Pair with Private and Shared Information]
\begin{rm}
Assume $\theta_0=-1.25$ and $\sigma_0=\sigma = 1$, $x_1$ and $x_2$ are private information seen by experts 1 and 2, respectively, $x_3$ is shared information, and $x_4$ is related to the binary event $y$. The binary event $y$ is 1 if $x_4>0$ and is 0 otherwise. Suppose expert~1 reports $p_1 = F_{n_1+n_s}(\tau_1) = F_2(-1.25)\approx 0.36$. Then we have the Bayesian ensemble depicted in Figure~\ref{Figure: Examples of Conjugate-Pair Bayesian Ensembles}d for all possible reports from the second expert. Compare this ensemble with private and shared information to Example~\ref{Example: Normal/Normal Pair}'s ensemble with private information only (Figure~\ref{Figure: Examples of Conjugate-Pair Bayesian Ensembles}c). The presence of shared information in Figure~\ref{Figure: Examples of Conjugate-Pair Bayesian Ensembles}d again reduces the degree of extremizing we see in Figure~\ref{Figure: Examples of Conjugate-Pair Bayesian Ensembles}c. The region of anti-extremizing enlarges as one piece of private information becomes shared information.\Halmos
\label{Example: Normal/Normal Pair with Private and Shared Information}
\end{rm}
\end{example}

For other settings of the conjugate-pair ensembles above, we get similar results. The ensembles tend to extremize the average forecast, although not always, even when there is no shared information. As more of the information is shared, the degree of extremizing reduces and the region of anti-extremizing enlarges. Anti-extremizing emerges (or increases) because of the reduction in the overall sample size the decision maker uses to aggregate the experts' forecast, as the experts share more information and see less information privately. The smaller sample size means the decision maker is less confident in his aggregate forecast.

\subsection{Comparison with Existing Aggregators}
The conjugate-pair ensembles above, because they are optimal aggregators under some reasonable assumptions, suggest an important structural property we would like to see in any aggregation method. An aggregation method should incorporate the prior-predictive probability $p_0$. The existing methods in Ranjan and Gneiting (2010)\nocite{ranj:gnei:2010}, Turner et al. (2014)\nocite{turn:stey:merk:bude:wall:2014}, Baron et al. (2014)\nocite{baro:mell:tetl:2014}, Satop\"{a}\"{a} et al. (2014), and Satop\"{a}\"{a} et al. (2016)\nocite{sato:pema:unga:2016}, however, assume $p_0$ is equal to one-half or do not incorporate $p_0$. In many applications though, such as in the two empirical studies of Section~\ref{Section: Empirical Studies}, $p_0$ (or the base rate of the $y$'s in a training set) will be far from one-half.

\begin{figure}[h!]
\begin{center}
\vspace{0cm}
\subfloat[KLOP and BLOP.]{\includegraphics[width=7cm]{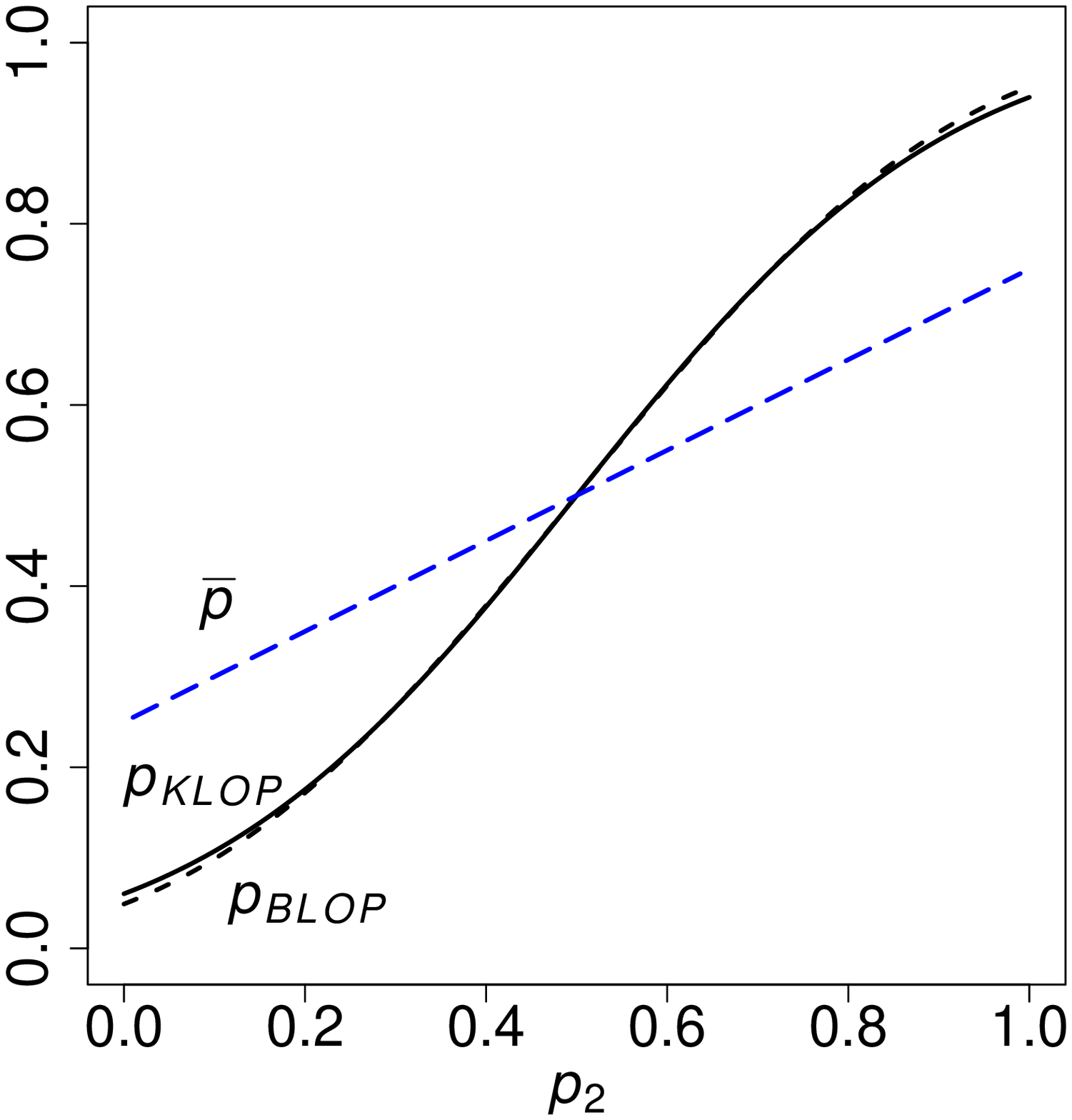}}
\subfloat[Probit Ensemble and Logit Aggregator.]{\includegraphics[width=7cm]{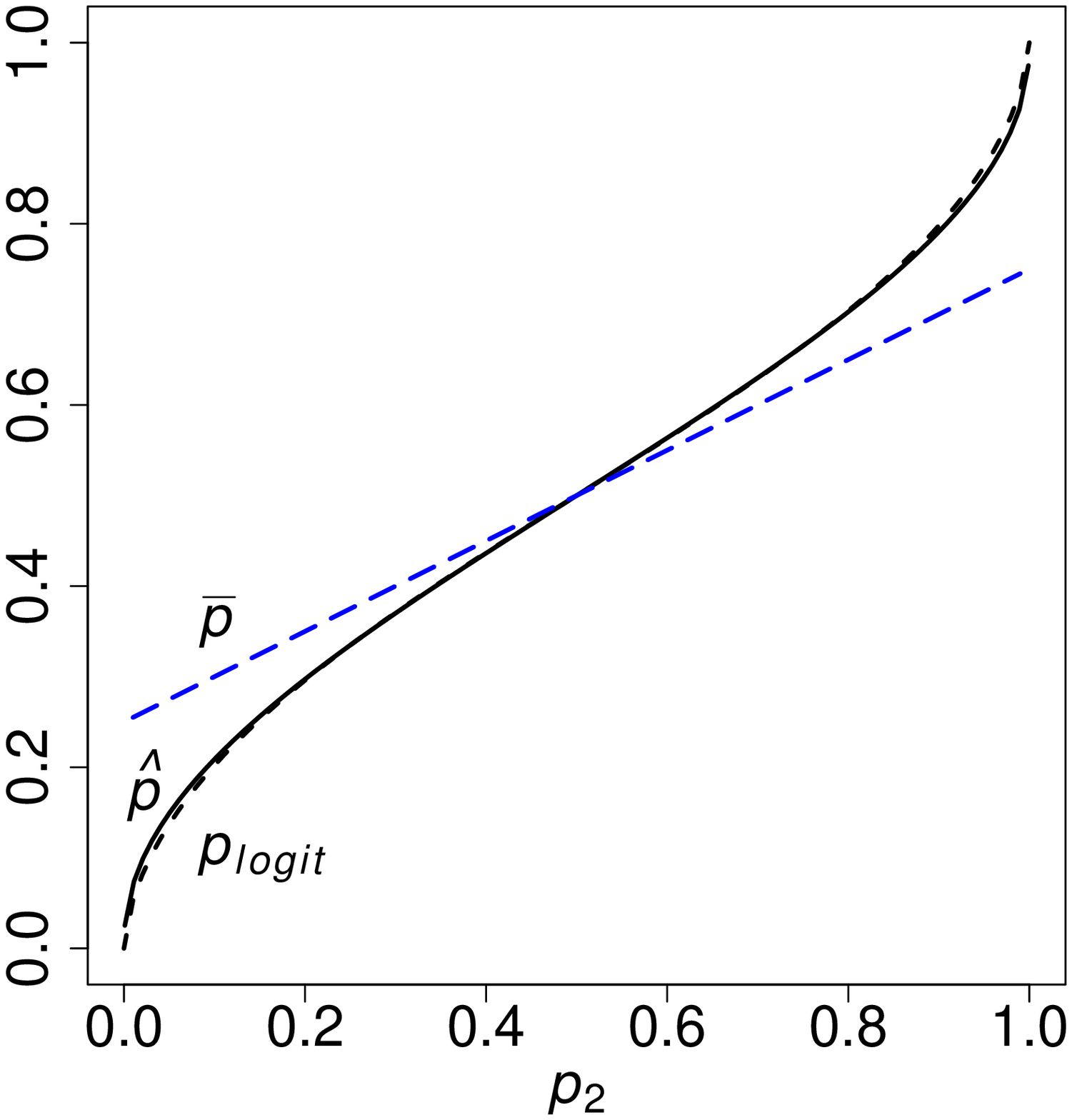}}
\vspace{1cm}
\caption{Ensembles With No Informative Prior-Predictive Probability Incorporated.}
\label{Figure: Examples of Other Ensembles}
\end{center}
\end{figure}

Ranjan and Gneiting (2010), Turner et al. (2014)\nocite{turn:stey:merk:bude:wall:2014}, and Baron et al. (2014)\nocite{baro:mell:tetl:2014} propose transformations of the linear opinion pool $p_{\mathit{LOP}} = \sum_{i=1}^k w_ip_i$ where $w_i > 0$ for $i=1,\ldots,k$, and $\sum_{i=1}^k w_i=1$. Ranjan and Gneiting (2010) study the beta-transformed linear opinion pool (BLOP): $p_{\mathit{BLOP}} = F_{\mathit{Be}(\alpha,\beta)}(p_{\mathit{LOP}})$ where $F_{\mathit{Be}(\alpha,\beta)}$ is the cdf of the beta distribution. Turner et al. (2014)\nocite{turn:stey:merk:bude:wall:2014} and Baron et al. (2014)\nocite{baro:mell:tetl:2014} use the Karmarkar equation $p_{\mathit{LOP}}^a/(p_{\mathit{LOP}}^a + (1-p_{\mathit{LOP}})^a)$ to extremize the linear opinion pool. We call this aggregator the Karmarkar-transformed linear opinion pool (KLOP), denoted by $p_{\mathit{KLOP}}$. Both methods involve s-shaped transformations of the linear opinion pool. These two methods, fit to the same data, will be almost identical. For example, it is difficult to distinguish the Karmarkar-transformed linear opinion pool with $a=2.5$ from the beta-transformed linear opinion pool with $\alpha=\beta=5$; see Figure~\ref{Figure: Examples of Other Ensembles}a where $p_{\mathit{LOP}}=\bar{p}$ and $p_1=1/2$.

Satop\"{a}\"{a} et al. (2014) and Turner et al. (2014)\nocite{turn:stey:merk:bude:wall:2014} propose the logit aggregator, a variant of the logarithmic opinion pool (French 1980, Dawid et al. 1995). This aggregator is a generalized additive model without an intercept: $p_{\mathit{logit}} = F_{\mathit{Lo}(0,1)}(\sum_{i=1}^{k} (a/k)F_{\mathit{Lo}(0,1)}^{-1}(p_i))$ where $F_{\mathit{Lo}(l,\psi)}(z) = 1/(1+e^{-(z-l)/\psi})$ is the cdf of the logistic distribution with location $l$ and scale $\psi$ and $F_{\mathit{Lo}(l,\psi)}^{-1}(p) = l + \psi\log(p/(1-p))$ is the inverse of its cdf. Ranjan and Gneiting (2010)\nocite{ranj:gnei:2010} and Satop\"{a}\"{a} et al. (2016) study versions of the probit ensemble in Example~\ref{Example: Normal/Normal Pair} with $p_0 = 1/2$. When $p_0 = 1/2$, the probit ensemble reduces to a generalized additive model with no intercept. These two ensembles are depicted in Figure~2b. The probit ensemble $\hat{p}$ in Figure~\ref{Figure: Examples of Other Ensembles}b has the same settings as the one in Figure~1c (except for $\theta_0 = 0$). This probit ensemble and the logit aggregator with $a=1.25$ are virtually indistinguishable.

Interestingly, both the logit aggregator and the probit ensemble in Figure~\ref{Figure: Examples of Other Ensembles}b involve inverse s-shaped functions. In the next section, we will see that negative correlation between experts' information states results in an s-shaped probit ensemble. The experts' information states according to the probit ensemble of Proposition~\ref{Proposition: Normal/Normal Pair Ensemble with Private and Shared Information} are always positively correlated (see $v_{i,j}$ in the Proof of Proposition~\ref{Proposition: Normal/Normal Pair Ensemble with Private and Shared Information}). Because negative correlation among experts is rare in practice, this result calls into question the use of s-shaped ensembles.

\section{Generalized Probit Ensemble}
\label{Section: Generalized Probit Ensemble}
Below we introduce the generalized probit ensemble. For this ensemble, we first assume the experts' information states (e.g., sufficient statistics) are jointly normally distributed, as in the probit ensemble of Example~\ref{Example: Normal/Normal Pair}, but with a general correlation structure. This general correlation structure can capture more complicated patterns of overlapping information sources, without having to work through detailed combinatorics (Clemen 1987\nocite{clem:1987}). Second, we assume a generalized linear model of these information states. This second assumption is informed by the Bayesian ensemble in Proposition~\ref{Proposition: Conjugate-Pair Ensemble with Private Information Only} and, in particular, by its generalized linear model form in \eqref{Equation: CP Bayesian Ensemble of Information States}.

At the outset, the decision maker places beliefs directly on the experts' information states---states that may result from experts seeing some observations in common. He assumes the experts' information states $\boldsymbol{x} = (x_1,\ldots,x_k)'$ are jointly normally distributed with mean $\boldsymbol{\mu}$ and covariance matrix $\boldsymbol{\Sigma}$, denoted $\boldsymbol{x} \sim N(\boldsymbol{\mu}, \boldsymbol{\Sigma})$. The covariance matrix has elements $\sigma_{ij}$, and the correlation between $x_i$ and $x_j$ is defined as $\rho_{ij} = \sigma_{ij}/(\sqrt{\sigma_{ii}}\sqrt{\sigma_{jj}})$.

The decision maker also assumes the conditional distribution of $y$ given the information states is given by the generalized linear model of the information states
\begin{equation}
\begin{split}
P(y=1|x_1,\ldots,x_k) = F_{z_0}(\alpha_0 + \alpha_1 x_1 + \cdots + \alpha_k x_k),
\label{Equation: Bayesian Ensemble of Information States}
\end{split}
\end{equation}
where the inverse link function $F_{z_0}$ is the cdf of a continuous random variable $z_0$ with the entire real line as its support. In addition, he assumes the inverse link function $F_{z_0}$ is the standard cdf from a location-scale family, i.e., the random variable $z = l + \psi z_0$ has the cdf $F_{z_0}((z-l)/\psi)$ where $l$ is the location parameter and $\psi$ is the scale parameter. Finally, he assumes any expert $i$ reports the probability $p_i = P(y=1|x_i)$.

In our generalized linear model, we do not restrict ourselves here to the standard normal link function in a generalized linear model of experts' information states. We propose a member of any location-scale family for the inverse link function and later focus on the exponential-power distribution. The exponential-power inverse link function has the flexibility to be either the standard normal near one extreme or the uniform on the other extreme. Therefore, the generalized probit ensemble can take the form of either an inverse s-shape or a linear shape, as needed.

\begin{proposition}[Generalized Probit Ensemble]
\begin{it}
Given the assumptions in this section, the optimal way to aggregate experts' forecasts is with the generalized probit ensemble
\begin{equation}
\begin{split}
\hat{p} = P(y=1|p_1,\ldots,p_k) = F_{z_0}\bigg(\beta_0 F_{z_0+\sqrt{v_0}x_0}^{-1}(p_0) + \sum_{i=1}^{k} \beta_i F_{z_0+\sqrt{v_i}x_0}^{-1}(p_i) \bigg),
\label{Equation: Generalized Probit Ensemble}
\end{split}
\end{equation}
where $p_0 = F_{z_0 +\sqrt{v_0} x_0}(m_0)$, $p_i = F_{z_0 +\sqrt{v_i} x_0}(m_0 + \beta_i^{-1}\alpha_i (x_i -\mu_i))$, $m_0 = \alpha_0 + \sum_{i=1}^{k}\alpha_i \mu_i$,
\begin{equation}
\begin{split}
\beta_0= 1-\sum_{i=1}^k \beta_i,  \,\, \mathit{and} \,\,
\beta_i= \frac{\alpha_i \sqrt{\sigma_{ii}}}{\sum_{j=1}^k \alpha_j \sqrt{\sigma_{jj}}\rho_{ij} } \,\, \mathit{for} \,\, i=1,\ldots,k.
\nonumber
\end{split}
\end{equation}
Also, the cdf $F_{z_0+\sqrt{v_i}x_0}$ is the cdf of the sum of the independent random variables $z_0$ and $\sqrt{v_i}x_0$ where $z_0$ is the standard random variable from a location-scale family, $x_0$ is a standard normal random variable,
\begin{equation}
\begin{split}
v_0 = \sum_{i =1}^k \sum_{j =1}^k \alpha_i \alpha_j \sigma_{ij}, \,\, \mathit{and} \,\, v_i = \sum_{j \neq i} \sum_{j' \neq i} \alpha_j \alpha_{j'} \sigma_{jj'} - \frac{\big(\sum_{j \neq i} \alpha_j \sigma_{ij}\big)^2}{\sigma_{ii}} \,\, \mathit{for} \,\, i=1,\ldots,k.
\nonumber
\end{split}
\end{equation}
\label{Proposition: Generalized Probit Ensemble}
\end{it}
\end{proposition}

Immediately we see that the generalized probit ensemble in \eqref{Equation: Generalized Probit Ensemble} is a generalized additive model of the experts' probabilities. The main benefit of this ensemble is that we do not need to work with a conjugate pair's predictive distribution, which can be difficult to do in all but few cases.

Another immediate consequence of the result is that each expert is calibrated. An expert is calibrated if $P(y=1|p_i) = p_i$ (French 1986\nocite{fren:1986}, Murphy and Winkler 1987\nocite{murp:wink:1987}). As was the case in the previous section, this calibration means that the decision maker, if he heard from only one expert, would use the same probability the expert reported. Note that if the expert is not calibrated, then the decision maker could re-calibrate the expert's reported probability using the re-calibration function $R_i(p_i)$ so that $P(y=1|x_i) = R(p_i)$. All results related to the generalized probit ensemble hold with $R(p_i)$ in place of $p_i$, if an expert is not calibrated.

To interpret the coefficients in the ensemble, it is helpful to consider the case of exchangeable experts. If the experts are exchangeable, then the weights are given by
\begin{equation}
\begin{split}
\beta_0= 1-\frac{k}{(k-1)\rho + 1} <0  \,\, \mathrm{and} \,\,
\beta_i= \frac{1}{(k-1)\rho + 1}>0 \,\, \mathrm{for} \,\, i=1,\ldots,k,
\nonumber
\end{split}
\end{equation}
Note that for the exchangeable distribution of information states to be a proper distribution, we must have $\rho > -1/(k-1)$, which ensures that the matrix $\boldsymbol{\Sigma}$ is positive definite.

\subsection{Exponential-Power Inverse Link Function}
Below we propose a family of inverse link functions to use when fitting the generalized probit ensemble to data. The family is based on the exponential-power distribution, also known as a generalized error or generalized Gaussian distribution (Subbotin 1923\nocite{subb:1923}, Box and Tiao 1973\nocite{box:tiao:1973}, Mineo and Ruggieri 2005\nocite{mine:rugg:2005}, Zhang et al. 2012\nocite{zhan:wang:liu:jord:2012}). This family contains the normal and Laplace distributions as special cases. Before we give the family's general form, we look at the example of the normal distribution.

\begin{example}[Normal Distribution]
\begin{rm}
Let the inverse link function $F_{z_0}$ in Proposition~4 be standard normal cdf. In this case, $F_{z_0+\sqrt{v_i}x_0}(u) = \Phi(u/\sqrt{1+v_i})$ so that $F_{z_0+\sqrt{v_i}x_0}^{-1}(p_i) = \sqrt{1+v_i}\Phi^{-1}(p_i)$. The probit ensemble, according to the assumptions in this section, is given by
\begin{equation}
\begin{split}
\hat{p} = \Phi\bigg(\beta_0\sqrt{1+v_0}\Phi^{-1}(p_0) + \sum_{i=1}^k \beta_i \sqrt{1+v_i}\Phi^{-1}(p_i)\bigg).
\label{Equation: Probit Ensemble II}
\nonumber
\end{split}
\end{equation}
\Halmos
\label{Example: Probit Ensemble II}
\end{rm}
\end{example}

To estimate this probit ensemble's coefficients $\beta_i' = \beta_i\sqrt{1+v_i}$ from data, one can fit a generalized linear model of $y$ on $(\Phi^{-1}(p_1),\ldots,\Phi^{-1}(p_k))$ using a standard normal inverse link function.

\begin{example}[Exponential-Power Distribution]
\begin{rm}
Let the inverse link function be the cdf $F_{z_0}$ with density
\begin{equation}
\begin{split}
f_{z_0}(z) = \frac{1}{2\eta^{1/\eta} \Gamma(1 + 1/\eta)\psi} \exp\bigg(-\frac{1}{\eta}\bigg|\frac{z-l}{\psi}\bigg|^{\eta}\bigg).
\nonumber
\end{split}
\end{equation}
This is the density of an exponential-power distribution. We denote this distribution by $z_0 \sim \mathit{EP}(l, \psi, \eta)$ where $l$ is the mean, $\eta^{2/\eta}\psi^2\Gamma(3/\eta)/\Gamma(1/\eta)$ is the variance, and $\eta >0$ is the power parameter. For a fixed power parameter, this distribution is a member of a location-scale family.\Halmos
\label{Example: Exponential-Power Distribution}
\end{rm}
\end{example}

For the ensemble based on the exponential-power distribution, the inside function is $F_{z_0 + \sqrt{v_i} x_0}$, which is not tractable (Soury and Alouini 2015\nocite{sour:alou:2015}). Nonetheless, we can approximate this inside function closely by matching the first two moments. We choose $v_i'$ in $\sqrt{1 + v_i'} z_{\mathit{EP}(0, 1, \eta)}$ so that the variance of this random variable equals the variance of $z_{\mathit{EP}(0, 1, \eta)} + \sqrt{v_i}x_0$. Their means are both zero by construction. Consequently, the approximate generalized probit ensemble with an exponential-power inverse link function, denoted by $\tilde{p}$, is given by
\begin{equation}
\begin{split}
\tilde{p} = F_{\mathit{EP}(0, 1, \eta)}\bigg(\beta_0\sqrt{1+v_0'}F_{\mathit{EP}(0, 1, \eta)}^{-1}(p_0) + \sum_{i=1}^k \beta_i \sqrt{1+v_i'}F_{\mathit{EP}(0, 1, \eta)}^{-1}(p_i)\bigg).
\label{Equation: Approximate Generalized Probit Ensemble}
\end{split}
\end{equation}
This ensemble is useful in applications because one can estimate the coefficients $\beta_i' = \beta_i\sqrt{1+v_i'}$ by fitting a generalized linear model of $y$ on $(F_{\mathit{EP}(0, 1, \eta)}^{-1}(p_1),\ldots,F_{\mathit{EP}(0, 1, \eta)}^{-1}(p_k))$ using the exponential-power inverse link function $F_{\mathit{EP}(0, 1, \eta)}$. For $\eta = 2$, the exponential-power distribution becomes the normal distribution, and the resulting ensemble is a probit ensemble. For $\eta = 1$, the exponential-power distribution becomes the Laplace distribution. For $\eta$ strictly less (greater) than 2, the distribution has fat (thin) tails. As $\eta \rightarrow \infty$, the distribution of $z_{\mathit{EP}(l, \psi, \eta)}$ goes to a uniform distribution on $(l-\psi, l+\psi)$ and the resulting ensemble approaches a linear ensemble, like the one in the beta/Bernoulli ensemble of Example~\ref{Example: Beta/Bernoulli Pair}.

\begin{figure}
\begin{center}
\vspace{0cm}
\subfloat[More Linear As $\eta$ Increases.]{\includegraphics[width=7cm]{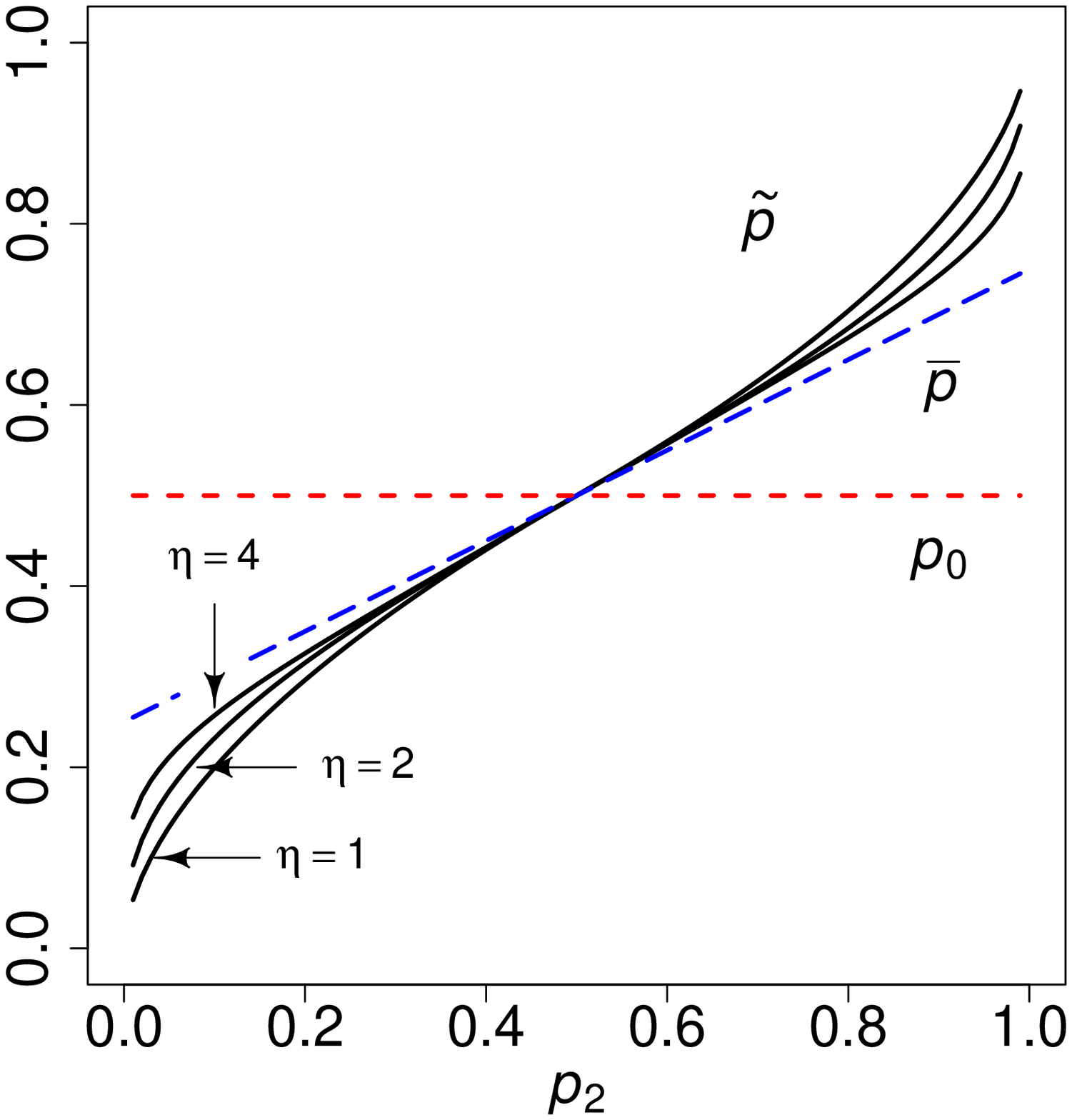}}
\subfloat[S-shaped with $\rho<0$.]{\includegraphics[width=7cm]{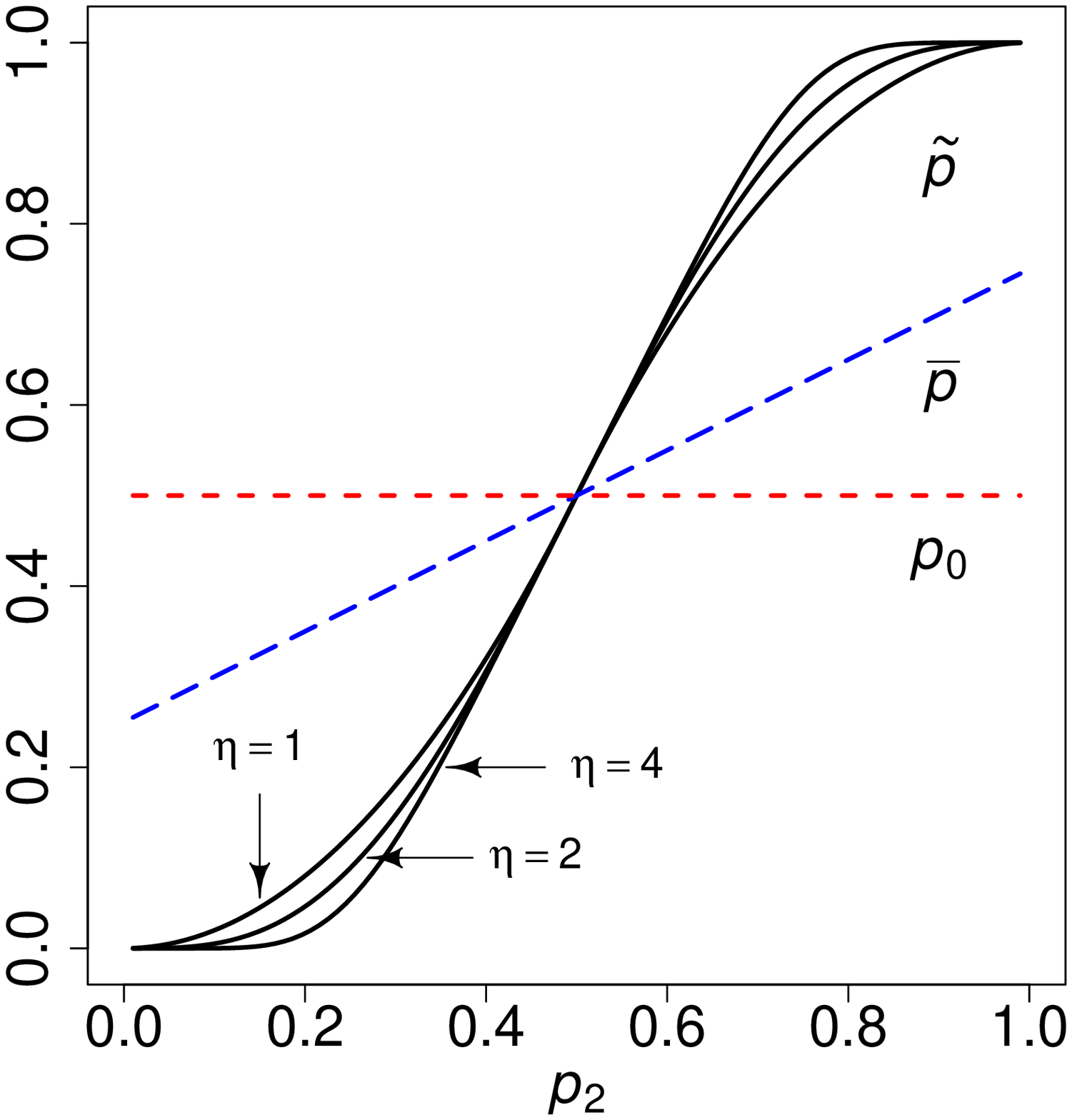}}
\vspace{1cm}
\caption{Approximate Generalized Probit Ensembles As the Power Parameter ($\eta$) and the Correlation Coefficient ($\rho$) Change.}
\label{Figure: Approximate Generalized Probit Ensembles}
\end{center}
\end{figure}

In Figure~\ref{Figure: Approximate Generalized Probit Ensembles}, we show several approximate generalized probit ensembles of two exchangeable experts' forecasts. We plot these ensembles as a function of $p_2$ with $p_1 = 1/2$, $\alpha =1$, $\sigma=1/20$, and $\rho = 3/4$. The weights on these positively correlated experts are $\beta_1 = \beta_2 = 0.57$. Figure~\ref{Figure: Approximate Generalized Probit Ensembles}a depicts three approximate generalized probit ensembles: the ensemble with $\eta = 1$, the probit ensemble ($\eta = 2$), and the ensemble with $\eta =4$.  As the power parameter increases, the ensemble becomes more linear. Also, we see in Figure~\ref{Figure: Approximate Generalized Probit Ensembles}b that the ensemble of positively correlated experts is inverse s-shaped, while the ensemble of negatively correlated ($\rho = -1/2$) experts is s-shaped.

\section{Empirical Studies}
\label{Section: Empirical Studies}
In this section, we present two empirical studies where we fit the generalized probit ensemble and compare its out-of-sample forecasting performance to several leading aggregation methods. The challenge in the first study is to predict defaults on loans acquired by Fannie Mae in 2007, just before the Great Recession.  These data are available at \url{https://loanperformancedata.fanniemae.com/lppub/index.html}. In the Fannie Mae data, there are 1,056,724 records on acquired loans and 20 independent variables, such as the borrower's credit score, the home's loan-to-value, and borrower's debt-to-income ratio. This year of acquisitions had the highest rate of defaults at 8.5\% in the period 2000-2015. The second study's challenge is to predict bad used-car buys by Carvana (a large used-car retailer) during the period 2009-2010. These data are available at \url{https://www.kaggle.com/c/DontGetKicked/data}. In the Carvana data, there are 72,983 records on used-car purchases and 34 independent variables, such as the vehicle's age, the vehicle's odometer, the vehicle's model, the buyer's id number, and the auction's location. This prediction challenge was part of a Kaggle competition called {\it Don't Get Kicked!}. The base rate of defective cars in the training set from Kaggle is 12.3\%.

\subsection{Training and Stacking the Models}
In the machine learning literature, forecast aggregation is known as stacking (Wolpert 1992\nocite{wolp:1992}, Breiman 1996\nocite{brei:1996}, Smyth and Wolpert 1999\nocite{smyt:wolp:1999}, D\v{z}eroski and \v{Z}enko 2004\nocite{dzer:zenk:2004}). The idea is that the predictions from several base models become features in a second-stage stacker model. Breiman (1996)\nocite{brei:1996} and Smyth and Wolpert (1999)\nocite{smyt:wolp:1999}, for example, both consider stacker models that are linear opinion pools of base models' probabilities. They choose optimal weights in a linear opinion pool that maximize the likelihood on a training set.

For each study, we trained three base models using the covariates available in the datasets. Each base model is a leading statistical or machine learning algorithm or part of a competition-winning ensemble. The first model is a regularized logistic regression (RLR), the lasso proposed by Tibshirani (1996)\nocite{tibs:1996}. The second model is the random forest (RF) introduced by Breiman (2001)\nocite{brei:2001}. The third model is the extreme gradient boosted trees model called xgboost (XGB) (Chen and Guestrin 2016)\nocite{chen:gues:2016}. This model is an extension of the gradient boosted trees model introduced by Friedman (2001)\nocite{frie:2001}. We could add more more base models to our ensemble, but our goal here is to demonstrate the plausibility of our approach.  The xgboost model, by itself, represents a difficult benchmark to beat. It was a part of 17 winning solutions published on Kaggle in 2015, and it was used by every team in the top 10 at KDD Cup 2015 (Chen and Guestrin 2016)\nocite{chen:gues:2016}.

To ensure that an ensemble was trained on out-of-sample probabilities from the base models, we employed a two-step process for building a stacker model. First, we randomly split the data into 10 equal folds and used these folds for both steps. In Step~1, the base models were fit to the first nine folds of data using the available covariates (e.g., credit score and loan-to-value) and the outcomes of the binary event of interest (e.g., loan defaults). Then the base models were used to predict the binary event of interest in the tenth fold.  Next the base models were trained on Folds~1-8 and 10 and used to predict the binary events of interest in the Fold~9. This process continued until each fold was held out once and out-of-sample predictions were made for it. In Step~2, an ensemble, or stacker model, was trained on the out-of-sample predictions made by the base models in Step~1. For example, a stacker model was trained on Folds~1-9 and tested (or evaluated) on Fold~10. This training and testing of a stacker model was done 10 times, with each fold serving as the hold-out sample once. For more details on how these models were trained and tested, see the supplemental materials. All data and code are available from the authors.

\subsection{Scoring the Models' Forecasts}
To evaluate out-of-sample forecasts, we use three different scoring rules: (i) the log score ($\mathit{LS}$), (ii) the asymmetric log score ($\mathit{ALS}$), and (iii) the area under the curve ($\mathit{AUC}$). The first scoring rule is negatively oriented (lower score is better), and the second and third scoring rules are positively oriented (higher score is better). The log score of a probability forecast $p$ for a binary event $y$ is given by $\mathit{LS}(p,y) = -(y\log(p)+(1-y)\log(1-p))$ for $0 < p <1$. This score is consistent with maximum likelihood estimation; the model (or ensemble) with the lowest average log score maximizes the log-likelihood. The asymmetric log score of a probability forecast $p$ for a binary event $y$ is given by $\mathit{ALS}(y,p) = (\mathit{LS}(c,y) - \mathit{LS}(p,y)) / \mathit{LS}(c, I_{p > c})$ where $I_{p > c}$ equals 1 if $p > c$ and equals 0 otherwise (Winkler 1994)\nocite{wink:1994}. This score is ``adjusted for the difficulty of the forecast task \ldots with the value of $c \in (0,1)$ adapted to reflect a baseline probability" (Gneiting and Raftery 2007, p. 365)\nocite{gnei:raft:2007}.  In the results we report below, $c$ is taken to be the base rate of occurrence of $y$ (denoted $\bar{y}$) in the training set. The area under the curve is a popular score in the machine-learning community (Hand and Till 2001)\nocite{hand:till:2001}.

\subsection{Results}
Table~\ref{Table: Average Scores} reports the average scores of out-of-sample predictions from the three base models ($p_{\mathit{RLR}}$, $p_{\mathit{RF}}$, and $p_{\mathit{XGB}}$), four existing aggregation models (the linear opinion pool with equal weights $\bar{p}$, the linear opinion pool with optimal weights $p_{\mathit{OLOP}}$, the beta-transformation of the linear opinion pool with optimal weights $p_{\mathit{BLOP}}$, the logit aggregator $p_{\mathit{logit}}$), and the best approximate generalized probit ensemble with an exponential-power inverse link function, $\tilde{p}$ in \eqref{Equation: Approximate Generalized Probit Ensemble}. The best $\tilde{p}$ has the power parameter $\eta^*$ that minimizes the average log loss out-of-sample. The estimates of $\eta^*$ are $40$ and $9$ in the two studies, respectively. In both studies, the best approximate generalized probit ensemble outperforms all other models on average over the 10 cross-validation folds.

\begin{table}[h!]
\centering
\begin{tabular}{lccccccc}
\hline
        &   \multicolumn{3}{c}{Fannie Mae}      & &   \multicolumn{3}{c}{Carvana}  \\
             \cline{2-4} \cline{6-8}
Model	&	$\mathit{LS}$	&	$\mathit{ALS}$	&	$\mathit{AUC}$	&	&	$\mathit{LS}$	&	$\mathit{ALS}$  &	$\mathit{AUC}$	\\
\hline
$p_{\mathit{RLR}}$	&	0.2362	&	0.2921	&	0.8162	&&	0.3109	&	0.1414	  &	0.7552	\\
$p_{\mathit{RF}}$	&	0.2450	&	0.2496	&	0.7940	&&	0.2993	&	0.1568	  &	0.7612	\\
$p_{\mathit{XGB}}$	&	0.2337	&	0.3058	&	0.8206	&&	0.2946	&	0.1687	 &	0.7708	\\
$\bar{p}$	        &	0.2347	&	0.2951	&	0.8205	&&	0.2966	&	0.1719	 &	0.7742	\\
$p_{\mathit{OLOP}}$	&	0.2330	&	0.3083	&	0.8229	&&	0.2932	&	0.1765	 &	0.7753	\\
$p_{\mathit{BLOP}}$	&	0.2359	&	0.2924	&	0.8229	&&	0.2953	&	0.1663	 &	0.7753	\\
$p_{\mathit{logit}}$ &	0.2329	&	0.3090	&	0.8229	&&	0.2928	&	0.1775	 &	0.7751	\\
$\tilde{p}$ 	&	{\bf 0.2327}	&	{\bf 0.3116}	&	{\bf 0.8230}	&&	{\bf 0.2925}	&	{\bf 0.1777}    &	{\bf 0.7753}	\\
\hline																								
\end{tabular}
\vspace{1cm}
\caption{Average Scores of Out-of-Sample Predictions in the Two Studies.} \label{Table: Average Scores}
\end{table}

After performing 10-fold cross validation for each study, we estimate the coefficients in the generalized linear model for the best approximate generalized probit ensemble on the entire dataset. Table~\ref{Table: Final Estimations of Best Generalized Probit Ensembles} lists these estimates. Not surprisingly, the ensemble puts the most weight on the best base model: xgboost. We also report in Table~\ref{Table: Final Estimations of Best Generalized Probit Ensembles} the base rate at which the binary events occur in each dataset.  In addition, we provide the percentage of times in out-of-sample forecasting that the best approximate generalized probit ensemble extremizes the average forecast.

\begin{table}[h!]
\centering
\begin{tabular}{llll}
\hline
                                                 &   Fannie Mae        &   Carvana            \\
\hline
Constant                                         &	 0.0496    &	0.0653    \\
Coefficient $\beta_{\mathit{RLR}}'$              &	0.3710	   &	-0.0281      \\
Coefficient $\beta_{\mathit{RF}}'$               &	0.0810     &	0.3864     \\
Coefficient $\beta_{\mathit{XGB}}'$              &	0.6070    &	0.7176     \\
\hline
Power parameter $\eta^*$	                     &	40 	               &	9	               \\
$\tilde{p}$ extremizes $\bar{p}$                   &  79.6\%             &    81.3\%             \\
\hline
Base rate $\bar{y}$                             &	0.0849	           &	0.1230	           \\
No. of observations	                             &	1,056,724 	       & 	72,983 	         \\
\hline
\end{tabular}
\vspace{1cm}
\caption{Final Estimation of Best Approximate Generalized Probit Ensembles.}
\label{Table: Final Estimations of Best Generalized Probit Ensembles}
\end{table}

\begin{figure}[h!]
\begin{center}
\subfloat[Study 1: Fannie Mae]{\includegraphics[width=7cm]{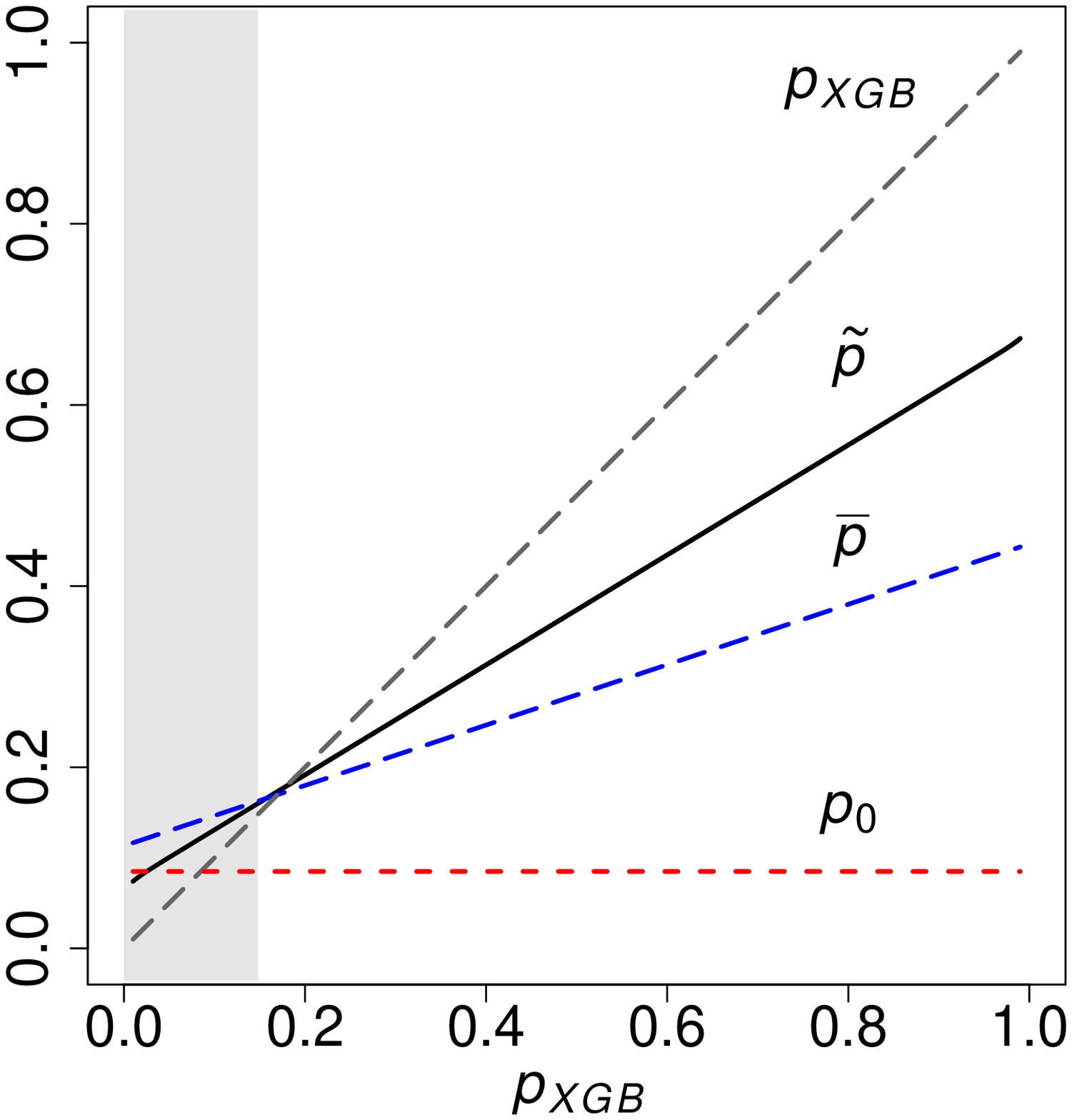}}
\subfloat[Study 2: Carvana]{\includegraphics[width=7cm]{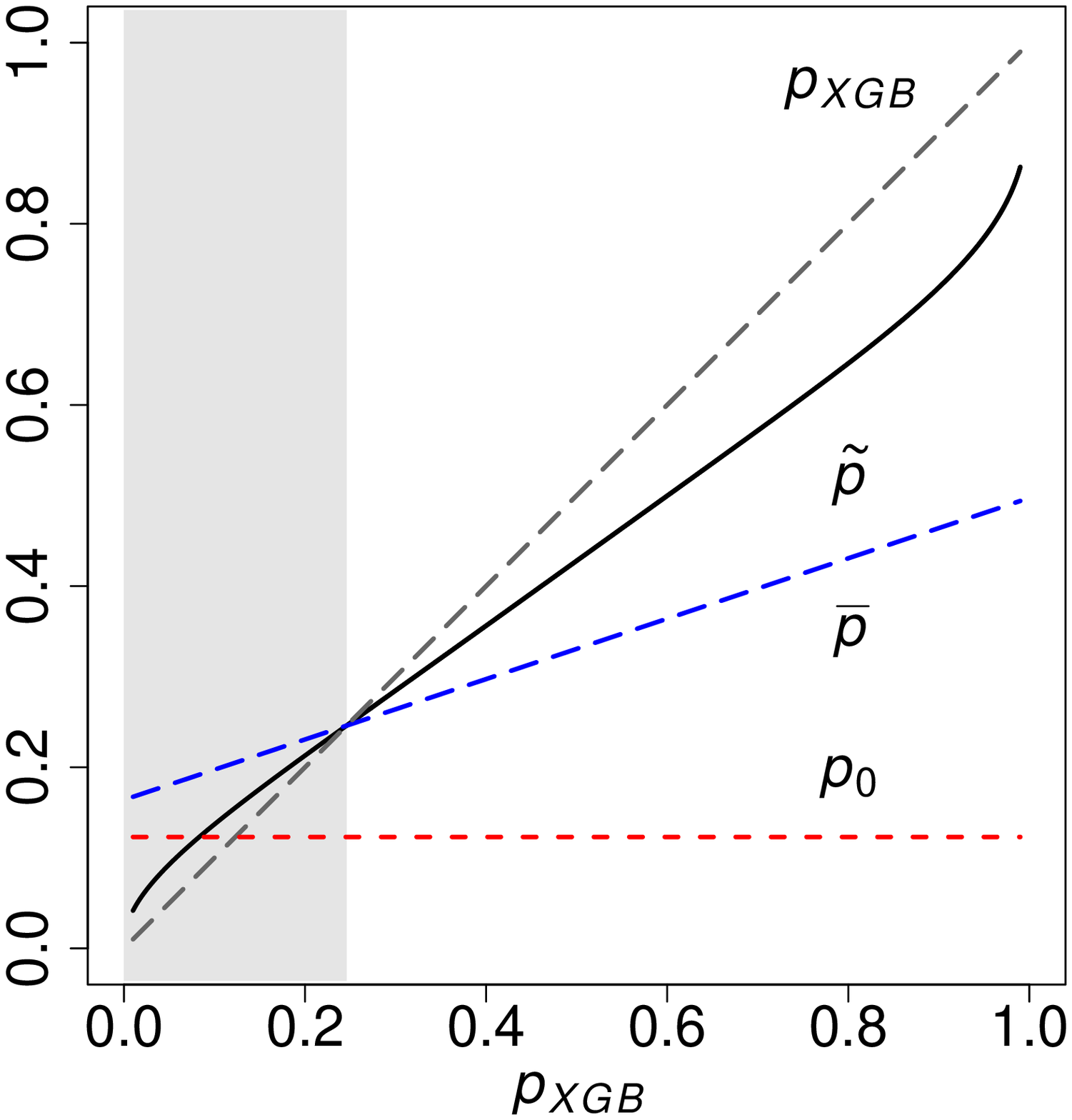}}
\vspace{1cm}
\caption{Best Approximate Generalized Probit Ensemble as a Function of the Best Base Model's Forecast.}
\label{Figure: Best Approximate Generalized Probit Ensemble}
\end{center}
\end{figure}

Figure~\ref{Figure: Best Approximate Generalized Probit Ensemble} depicts the best approximate generalized probit ensemble as a function of the best base model's forecasts, with the other two base models' forecasts set to twice the base rate in their respective dataset. For these settings, the ensembles lie somewhere in between all weight on the best base model (xgboost) and equal weights on the base models. In each plot, we also highlight the anti-extremizing region in gray. In addition, because the estimated power parameters are high in these studies, we can see that the ensembles are nearly linear over much of the domain of $p_{\mathit{XGB}}$.

\section{Summary and Conclusions}
\label{Section: Summary and Conclusions}
In this paper, we introduce a large class of Bayesian ensembles. Because the ensembles in this class are based on Bayesian reasoning, they can help us understand why, when, and how much extremizing is appropriate. It is appropriate to extremize the average forecast in order to remove the redundant reports of the prior-predictive probability in the experts' forecasts. When some information is shared by the experts, however, the decision maker may sometimes want to anti-extremize. Due to the shared information, the decision maker does not have as much total information upon which to base his aggregate forecast, and he may in turn issue a less confident aggregate forecast. This theoretical result matches the empirical findings of Tetlock and Gardner (2015\nocite{tetl:gard:2016}). They also find that when teams are good at sharing information, like teams of superforecasters, extremizing does not help them much.

The two types of Bayesian ensembles we introduce here---conjugate-pair Bayesian ensembles and the generalized probit ensemble---take the form of a generalized additive model. In the first type, the link function is determined by a regular, one-parameter exponential family and its conjugate prior.  This type is easy to interpret because of the natural sampling process that underlies the ensemble. In the second type, the inverse link function is any distribution from a location-scale family. This type is more flexible; and therefore more suitable for applications. In practice, we find that the exponential-power distribution is a good choice for the inverse link function in the generalized probit ensemble. Within the family of generalized probit ensembles are the probit ensemble near one extreme and a linear ensemble at the other extreme, depending on where the power parameter is set. Using a generalized linear modeling framework, this power parameter can be tuned to fit real world training data. Importantly, our ensembles include a constant term inside their generalized additive model forms. This constant term incorporates in the aggregate forecast the prior-predictive probability, e.g., the base rate of occurrences of the binary event in a training set.

In our two empirical studies, the generalized probit ensemble performed well in making out-of-sample predictions. It outperformed three base models and four leading aggregation methods. The base models in these studies were leading statistical and machine learning algorithms: the lasso, random forest, and xgboost. The latter two models are ensembles themselves and can be difficult to beat in practice, especially xgboost. In outperforming xgboost, it is important to stress that the Bayesian ensembles here are not replacements for an ensemble like xgboost. The Bayesian ensemble we fit used xgboost as one of its inputs. More importantly, the empirical results presented here demonstrate the plausibility of our generalized probit ensemble, not its superiority. We see several possible avenues for future work, including applications of the generalized probit ensemble to (possibly poorly calibrated) human judgments.

\section*{Appendix}
In this appendix, we provide proofs of Propositions~1-4 and derivations of Examples~1-4. We also state a lemma and its proof. This lemma is useful in the proof of Proposition~4.

\subsection*{Proof of Proposition~\ref{Proposition: Conjugate-Pair Ensemble with Private Information Only}.}
By Bernardo and Smith (2000, Prop. 5.5)\nocite{bern:smit:2000}, expert $i$'s forecast of $x$ given $\boldsymbol{x}_i$ is given by
\begin{equation}
\begin{split}
f(x|\boldsymbol{x}_i) = a(x)\frac{K(\tau_0 + n_i + 1, \tau_1 + t_i + h(x))}{K(\tau_0 + n_i, \tau_1 + t_i)},
\nonumber
\label{Equation: Expert posterior-predictive}
\end{split}
\end{equation}
where $t_i = \sum_{j = N_{i-1} + 1}^{N_i} h(x_j)$ is the sufficient statistic for $\boldsymbol{x}_i$. Expert $i$'s forecast of $(y|\boldsymbol{x}_i)$ is given by $P(y =1 | \boldsymbol{x}_i) = \int_{x \in A} f(x|\boldsymbol{x}_i)\, dx$.  Hence, $p_i = P(y =1 | \boldsymbol{x}_i) = F_{n_i}(\tau_1 + t_i)$. The decision maker, if he had direct access to the experts' samples, would form the following forecast of $(y|\boldsymbol{x}_1,\ldots,\boldsymbol{x}_k)$:
\begin{equation}
\begin{split}
P(y=1|\boldsymbol{x}_1,\ldots,\boldsymbol{x}_k) &= \int_{x \in A} a(x)\frac{K(\tau_0 + \sum_{i=1}^k n_i + 1, \tau_1 + \sum_{i =1}^k t_i + h(x))}{K(\tau_0 + \sum_{i=1}^k n_i, \tau_1 + \sum_{i =1}^k t_i)} \, dx \\
                                              &= F_{N_k}\bigg(\tau_1 + \sum_{i =1}^k t_i\bigg).
\label{Equation: DM posterior-predictive}
\end{split}
\end{equation}
The decision maker knows all the sample sizes because of the common knowledge assumption. The prior-predictive probability is given by
\begin{equation}
\begin{split}
p_0 = P(y=1) = \int_{x \in A} a(x)\frac{K(\tau_0 + 1, \tau_1 + h(x))}{K(\tau_0, \tau_1)} \, dx = F_{0}(\tau_1).
\nonumber
\label{Equation: Prior probability}
\end{split}
\end{equation}
By the assumption that $F_n(t)$ is strictly monotonic in $t$, we can invert $F_{n}(t)$. By substitution according to $\tau_1 = F_0^{-1}(p_0)$ and $t_i = F_{n_i}^{-1}(p_i)-\tau_1$ into \eqref{Equation: DM posterior-predictive}, we have the result.

\subsection*{Derivation of Example~\ref{Example: Beta/Bernoulli Pair}'s Ensemble.}  The Bernoulli distribution is a regular, one-parameter exponential family with pmf
\begin{equation}
\begin{split}
f(x_j|\theta) &= \theta^{x_j} (1-\theta)^{1-x_j} = (1-\theta)\exp\bigg(\log\bigg(\frac{\theta}{1-\theta}\bigg)x_j\bigg)  \ \mathrm{for}\  x_j \in \{0,1\},
\nonumber
\label{Equation: Bernoulli Likelihood}
\nonumber
\end{split}
\end{equation}
where $a(x_j) = 1$, $b(\theta) = 1-\theta$, $c(\theta) = \log(\frac{\theta}{1-\theta})$, and $h(x_j) = x_j$. Its conjugate prior has the pdf
\begin{equation}
\begin{split}
f(\theta) &= \frac{\Gamma(\alpha + \beta)}{\Gamma(\alpha)\Gamma(\beta)} \theta^{\alpha-1} (1-\theta)^{\beta-1} = \frac{1}{K(\tau_0,\tau_1)} (1-\theta)^{\tau_0} \exp\bigg(\log\bigg(\frac{\theta}{1-\theta}\bigg)\tau_1\bigg) \ \mathrm{for}\  \theta \in (0,1),
\label{Equation: Beta Prior for Bernoulli Likelihood}
\nonumber
\end{split}
\end{equation}
where $\tau_0 = \alpha + \beta - 2$, $\tau_1 = \alpha - 1$, and $K(\tau_0, \tau_1) = \Gamma(\tau_1 + 1)\Gamma(\tau_0 - \tau_1 + 1)/\Gamma(\tau_0 + 2)$. With $A = \{1\}$,
\begin{equation}
\begin{split}
F_{n}(t) &= \int_{x \in A} a(x)\frac{K(\tau_0 + n + 1, t + h(x))}{K(\tau_0 + n, t)} \, dx \\
         &=   \frac{\Gamma(t + 1 + 1)\Gamma(\tau_0 + n + 1 - (t + 1) + 1)}{\Gamma(\tau_0 + n + 1 + 2)}  \frac{\Gamma(\tau_0 + n + 2)}{\Gamma(t + 1)\Gamma(\tau_0 + n - t + 1)}  = \frac{t+1}{\alpha + \beta + n},
\label{Equation: Bernoulli Inverse Link}
\nonumber
\end{split}
\end{equation}
and $F_{n_i}^{-1}(p_i) = (\alpha + \beta + n_i)p_i - 1$.

\subsection*{Derivation of Example~\ref{Example: Gamma/Poisson Pair}'s Ensemble.}   The Poisson distribution is a regular, one-parameter exponential family with pmf
\begin{equation}
\begin{split}
f(x_j|\theta) &= \frac{\theta^{x_j} \exp(-\theta) }{x_j!}= \frac{1}{x_j!} \exp(-\theta)  \exp(\log(\theta) x_j)  \ \mathrm{for}\  x_j \in \{0,1,2,\ldots\},
\label{Equation: Poisson Likelihood}
\nonumber
\end{split}
\end{equation}
where $a(x_j) = 1/x_j!$, $b(\theta) = \exp(-\theta)$, $c(\theta) = \log(\theta)$, and $h(x_j) = x_j$. Its conjugate prior has the pdf
\begin{equation}
\begin{split}
f(\theta) = \frac{\beta^{\alpha}}{\Gamma(\alpha)} \theta^{\alpha-1} \exp(-\beta\theta) = \frac{1}{K(\tau_0,\tau_1)} \theta^{\tau_1} \exp(-\tau_0\theta) \ \mathrm{for}\  \theta \in (0,\infty),
\label{Equation: Gamma Prior for Possion Likelihood}
\nonumber
\end{split}
\end{equation}
where $\tau_0 = \beta$, $\tau_1 = \alpha -1$, and $K(\tau_0, \tau_1) = \Gamma(\tau_1 + 1)/\tau_0^{\tau_1 + 1}$. With $A = \{0\}$,
\begin{equation}
\begin{split}
F_{n}(t) &= \int_{x \in A} a(x)\frac{K(\tau_0 + n + 1, t + h(x))}{K(\tau_0 + n, t)} \, dx =   \frac{\Gamma(t + 1)}{(\tau_0 + n + 1)^{t+1}}     \frac{(\tau_0 + n)^{t + 1}}{\Gamma(t + 1)}  =  \exp( v_{n}(t + 1)),
\label{Equation: Poisson Inverse Link}
\nonumber
\end{split}
\end{equation}
where $v_{n} = \log((\beta + n)/(\beta + n + 1))$, and $F_{n_i}^{-1}(p_i) = \log(p_i)/v_{n_i} -1$.

\subsection*{Derivation of Example~\ref{Example: Normal/Normal Pair}'s Ensemble.} This example's normal distribution is a regular, one-parameter exponential family with pdf
\begin{equation}
\begin{split}
f(x_j|\theta) &= \frac{1}{\sqrt{2 \pi \sigma^2}}\exp\bigg(-\frac{1}{2\sigma^2}(x_j-\theta)^2\bigg) \\
              &= \frac{1}{\sqrt{2 \pi \sigma^2}}\exp\bigg(-\frac{x_j^2}{2\sigma^2}\bigg)\exp\bigg(-\frac{\theta^2}{2\sigma^2}\bigg)\exp\bigg(\frac{\theta}{\sigma^2}x_j\bigg) \ \mathrm{for}\  x_j \in (-\infty,\infty),
\label{Equation: Normal Likelihood}
\nonumber
\end{split}
\end{equation}
where $a(x) = (1/\sqrt{2 \pi \sigma^2})\exp(-x_j^2/(2\sigma^2))$, $b(\theta) = \exp\big(-\frac{\theta^2}{2\sigma^2}\big)$, $c(\theta) = \theta$, and $h(x_j) = x_j$. Its conjugate prior has the pdf
\begin{equation}
\begin{split}
f(\theta) &= \frac{1}{\sqrt{2 \pi \sigma_0^2}}\exp\bigg(-\frac{1}{2\sigma_0^2}(\theta-\theta_0)^2\bigg) \\
          &= \frac{1}{\sqrt{2 \pi \sigma_0^2}}\exp\bigg(-\frac{\theta_0^2}{2\sigma_0^2}\bigg) \bigg(\exp\bigg(-\frac{\theta^2}{2\sigma^2}\bigg) \bigg)^{\tau_0} \exp\bigg(\frac{\theta}{\sigma^2}\tau_1\bigg) \ \mathrm{for}\  \theta \in (-\infty,\infty),
\label{Equation: Normal Prior for Normal Likelihood}
\nonumber
\end{split}
\end{equation}
where $\tau_0 = \sigma^2/\sigma_0^2$, $\tau_1 = \sigma^2\theta_0/\sigma_0^2$, and $K(\tau_0, \tau_1) = \sqrt{(2 \pi \sigma^2)/\tau_0}\exp(\tau_1^2/(2\sigma^2\tau_0))$. With $A = (0,\infty)$,
\begin{equation}
\begin{split}
F_{n}(t) &= \int_{x \in A} a(x)\frac{K(\tau_0 + n + 1, t + h(x))}{K(\tau_0 + n, t)} \, dx \notag \\
         &= \int_0^{\infty}  \frac{1}{\sqrt{2\pi\frac{\tau_0+n +1}{\tau_0+n}\sigma^2}} \exp\bigg(-\frac{\tau_0+n}{2\sigma^2(\tau_0+n+1)} \bigg(x-\frac{t}{\tau_0+n}\bigg)^2\bigg) \, dx  \label{Equation: Normal Inverse Link} \\
         &= 1-\Phi\bigg(\frac{-\frac{t}{\tau_0+n}}{\sqrt{\frac{\tau_0+n +1}{\tau_0+n}\sigma^2}}\bigg)  = \Phi\bigg(\frac{\frac{t}{\tau_0+n}}{\sqrt{\frac{\tau_0+n +1}{\tau_0+n}\sigma^2}}\bigg)= \Phi\bigg(\frac{t}{\sqrt{v_n}} \bigg),
\nonumber
\end{split}
\end{equation}
where $v_n = (\sigma^2/\sigma_0^2+n)(\sigma^2/\sigma_0^2+n +1)\sigma^2$, and $F_{n_i}^{-1}(p_i) = \sqrt{v_{n_i}}\Phi^{-1}(p_i)$. The second integrand in \eqref{Equation: Normal Inverse Link} is the posterior-predictive distribution for the normal/normal model with known precision in Bernardo and Smith (2000, p. 439). For example, with $t=\tau_1 + t_i$ and $n = n_i$, this integrand is expert $i$'s posterior-predictive distribution of $x$.

\subsection*{Derivation of Example~\ref{Example: Generalized-Gamma/Gumbel Pair}'s Ensemble.}  This example's Gumbel distribution is a regular, one-parameter exponential family with pdf
\begin{equation}
\begin{split}
f(x_j|\theta) &= \frac{1}{\sigma}\exp\bigg(-\frac{x_j-\theta}{\sigma}\bigg) \exp\bigg(-\exp\bigg(-\frac{x_j-\theta}{\sigma}\bigg)\bigg) \\
              &= \frac{1}{\sigma}\exp\bigg(-\frac{x_j}{\sigma}\bigg) \exp\bigg(\frac{\theta}{\sigma}\bigg)  \exp\bigg(-\exp\bigg(-\frac{x_j-\theta}{\sigma}\bigg)\bigg) \ \mathrm{for}\  x_j \in (-\infty,\infty),
\label{Equation: Gumbel Likelihood}
\nonumber
\end{split}
\end{equation}
where $a(x_j) = \exp(-x_j/\sigma)/\sigma$, $b(\theta) = \exp(\theta/\sigma)$, $c(\theta) = -\exp(\theta/\sigma)$, and $h(x_j) = \exp(-x_j/\sigma)$. Its conjugate prior has the pdf
\begin{equation}
\begin{split}
f(\theta) = \frac{\beta^{\alpha}}{\sigma\Gamma(\alpha)}\bigg(\exp\bigg(\frac{\theta}{\sigma}\bigg)\bigg)^{\alpha} \exp\bigg(-\beta \exp\bigg(\frac{\theta}{\sigma}\bigg)\bigg) \ \mathrm{for}\  \theta \in (-\infty,\infty),
\label{Equation: Prior for Gumbel Likelihood}
\nonumber
\end{split}
\end{equation}
where $\tau_0 = \alpha$, $\tau_1 = \beta$, and $K(\tau_0,\tau_1) = \Gamma(\tau_0)\tau_1^{-\tau_0}\sigma$. With $A = (-\infty,0)$,
\begin{equation}
\begin{split}
F_{n}(t) &= \int_{x \in A} a(x)\frac{K(\tau_0 + n + 1, t + h(x))}{K(\tau_0 + n, t)} \, dx \\
         &= \int_{-\infty}^0 \frac{1}{\sigma}\exp\bigg(-\frac{x}{\sigma}\bigg)\frac{\Gamma(\alpha + n + 1) \big(t + \exp\big(-\frac{x}{\sigma}\big)\big)^{-(\alpha + n + 1)}}{\Gamma(\alpha + n) t^{-(\alpha + n)}} \, dx \\
         &= \frac{1}{\sigma}t^{\alpha + n} \int_{-\infty}^0 \frac{(\alpha + n)\exp\big(-\frac{x}{\sigma}\big)}{\big(t + \exp\big(-\frac{x}{\sigma}\big)\big)^{\alpha + n + 1}} \, dx =  t^{\alpha + n} \bigg[  \frac{1}{\big(t + \exp\big(-\frac{x}{\sigma}\big)\big)^{\alpha + n}} \bigg]_{-\infty}^0 = \bigg(\frac{t}{1 + t}\bigg)^{\alpha + n},
\label{Equation: Gumbel Inverse Link}
\nonumber
\end{split}
\end{equation}
and $F_{n_i}^{-1}(p_i) = p_i^{v_{n_i}}/(1-p_i^{v_{n_i}})$, where $v_{n} = 1/(\alpha + n)$.

\subsection*{Proof of Proposition~\ref{Proposition: Conjugate-Pair Ensemble with Private and Shared Information}.}
By Bernardo and Smith (2000, Prop. 5.5)\nocite{bern:smit:2000}, expert $i$'s forecast of $x$ given the private sample $\boldsymbol{x}_i$ and the shared sample $\boldsymbol{x}_{s}$ is given by
\begin{equation}
\begin{split}
f(x|\boldsymbol{x}_i,\boldsymbol{x}_{s}) = a(x)\frac{K(\tau_0 + n_i + n_{s} + 1, \tau_1 + t_i + t_{s} + h(x))}{K(\tau_0 + n_i + n_{s}, \tau_1 + t_i + t_{s})},
\label{Equation: Expert posterior-predictive II}
\nonumber
\end{split}
\end{equation}
where $t_i$ and $t_s$ are the sufficient statistics for the samples $\boldsymbol{x}_i$ and $\boldsymbol{x}_s$, respectively. Expert $i$'s forecast of $(y|\boldsymbol{x}_i,\boldsymbol{x}_{s})$ is given by $P(y =1 | \boldsymbol{x}_i,\boldsymbol{x}_{s}) = \int_{x \in A} f(x|\boldsymbol{x}_i,\boldsymbol{x}_{s})\, dx$.  Hence, $p_i = P(y =1 | \boldsymbol{x}_i,\boldsymbol{x}_{s}) = F_{n_i+n_{s}}(\tau_1 + t_i+ t_{s})$. The decision maker, if he had direct access to the experts' private samples and the shared sample, would form the following forecast of $(y|\boldsymbol{x}_1,\ldots,\boldsymbol{x}_k,\boldsymbol{x}_{s})$:
\begin{equation}
\begin{split}
P(y=1 |\boldsymbol{x}_1,\ldots,\boldsymbol{x}_k,\boldsymbol{x}_{s}) &= \int_{x \in A} a(x)\frac{K(\tau_0 + \sum_{i=1}^{k} n_i + n_s + 1, \tau_1 + \sum_{i =1}^k t_i + t_{s} + h(x))}{K(\tau_0 + \sum_{i=1}^{k} n_i + n_s, \tau_1 + \sum_{i =1}^k t_i + t_{s})} \, dx \\
&= F_{N_{k}+n_s}\bigg(\tau_1 + \sum_{i =1}^k t_i + t_{s}\bigg).
\label{Equation: DM posterior-predictive II}
\end{split}
\end{equation}
The decision maker knows all the sample sizes because of the common knowledge assumption. From the proof of Proposition~1, the prior-predictive probability is given by $p_0 = P(y=1) = F_{0}(\tau_1)$. By the assumption that $F_n(t)$ is strictly monotonic in $t$, we can invert $F_{n}(t)$. By substitution according to $\tau_1 = F_0^{-1}(p_0)$ and $t_i = F_{n_i+n_{s}}^{-1}(p_i)-\tau_1 - t_{s}$ into \eqref{Equation: DM posterior-predictive II}, we have
\begin{equation}
\begin{split}
P(y=1 |t_1,\ldots,t_k,t_{s}) &= F_{N_{k}+n_s}\bigg(\tau_1 + \sum_{i =1}^k t_i + t_{s}\bigg) \\
                            &= F_{N_{k}+n_s}\bigg(F_0^{-1}(p_0) + \sum_{i =1}^k (F_{n_i+n_{s}}^{-1}(p_i)-F_0^{-1}(p_0) - t_{s}) + t_{s}\bigg) \\
                            &= F_{N_{k}+n_s}\bigg(-(k-1) (F_0^{-1}(p_0) + t_{s}) + \sum_{i =1}^k F_{n_i+n_{s}}^{-1}(p_i)\bigg) \\
                            &= P(y=1|p_1,\ldots,p_k,t_{s}).
\nonumber
\end{split}
\end{equation}
Consequently, because $P(y=1|p_1,\ldots,p_k) = E[E(y|p_1,\ldots,p_k,t_{s})]$, we have the result.

\subsection*{Proof of Proposition~\ref{Proposition: Normal/Normal Pair Ensemble with Private and Shared Information}}
The first step in the proof is to derive $f(t_{s}|p_1,\ldots,p_k)$, which is equivalent to $f(t_{s}|t_1 + t_{s},\ldots,t_k + t_{s})$ because $t_i + t_{s} = \sqrt{v_{n_i+n_{s}}}\Phi^{-1}(p_i)-\tau_1$ is a known function of $p_i$. The fact that $(t_{s}, t_1 + t_{s},\ldots,t_k + t_{s})$ is jointly normally distributed follows from two facts: (a) $(x_1,\ldots,x_{N_{k}+n_s})$ is jointly normally distributed according to Bernardo and Smith's (2000, Prop.~5.5(ii)) and (b) $(t_{s}, t_1 + t_{s},\ldots,t_k + t_{s})$ is a linear combination of $(x_1,\ldots,x_{N_{k}+n_s})$. The requisite means are
\begin{equation}
\begin{split}
m_{s} = E[t_{s}] &= E\bigg[E\bigg[\sum_{j=N_k+1}^{N_{k}+n_s} x_j\bigg|\theta\bigg]\bigg] = E[n_{s} \theta] =  n_{s} \theta_0 \\
m_{i} = E[t_i + t_{s}] &= E\bigg[E\bigg[\sum_{j=N_{i-1}+1}^{N_i} x_j + \sum_{j=N_k+1}^{N_{k}+n_s} x_j\bigg|\theta\bigg]\bigg] = E[(n_i+n_{s}) \theta] =  (n_i+n_{s})\theta_0.
\nonumber
\end{split}
\end{equation}
The variance of $t_i + t_{s}$ is given by
\begin{equation}
\begin{split}
v_{i,i} &= \mathit{Var}[t_i + t_{s}] = E[\mathit{Var}[t_i + t_{s}|\theta]] + \mathit{Var}[E[t_i + t_{s}|\theta]] \\
&= (n_i+n_{s})\sigma^2 +  \mathit{Var}[(n_i+ n_{s})\theta] = (n_i+n_{s})\sigma^2 +  (n_i+ n_{s})^2\sigma_0^2.
\nonumber
\end{split}
\end{equation}
Similarly, the variance of $t_{s}$ is given by $v_{s,s} = \mathit{Var}[t_{s}] = n_{s}\sigma^2 +  n_{s}^2\sigma_0^2$. The covariance between $t_i + t_{s}$ and $t_j + t_{s}$ (for $i \neq j$) is given by
\begin{equation}
\begin{split}
v_{i,j} &= \mathit{Cov}[t_i + t_{s},t_j + t_{s}] = E[\mathit{Cov}[t_i + t_{s},t_j + t_{s}|\theta]] + \mathit{Cov}[E[t_i + t_{s}|\theta],E[t_j + t_{s}|\theta]] \\
&= E[\mathit{Cov}[t_i,t_j|\theta] + \mathit{Cov}[t_i,t_{s}|\theta] + \mathit{Cov}[t_{s},t_j|\theta] + \mathit{Cov}[t_{s},t_{s}|\theta]] \\
&\phantom{=} + \mathit{Cov}[E[t_i|\theta] + E[t_{s}|\theta],E[t_j|\theta] + E[t_{s}|\theta]] \\
&= n_{s}\sigma^2 + \mathit{Cov}[n_i\theta,n_j\theta] + \mathit{Cov}[n_i\theta,n_{s}\theta] + \mathit{Cov}[n_{s}\theta,n_j\theta] + \mathit{Cov}[n_{s}\theta,n_{s}\theta] \\
&= n_{s}\sigma^2 + (n_i n_j + n_i n_{s} + n_{s}n_j +n_{s}^2)\sigma_0^2,
\nonumber
\end{split}
\end{equation}
which follows by the law of total covariance and the fact that $t_i$ and $t_j$ are conditionally independent given $\theta$. Similarly, the covariance between $t_{s}$ and $t_j + t_{s}$ is given by $v_{s,j} = \mathit{Cov}[t_{s},t_j + t_{s}] = n_{s}\sigma^2 + (n_{s}n_j +n_{s}^2)\sigma_0^2$.

Thus, $(t_{s}, t_1 + t_{s},\ldots,t_k + t_{s})' \sim N(\boldsymbol{m}, \boldsymbol{V})$ (West and Harrison 1997, p. 637). The mean vector is $\boldsymbol{m} = (\boldsymbol{m}_1, \boldsymbol{m}_2)'$ where $\boldsymbol{m}_1 = m_{s}$ and $\boldsymbol{m}_2 = (m_1, \ldots, m_k)'$. The covariance matrix is
\begin{equation}
\begin{split}
\boldsymbol{V} = \bigg(
                   \begin{array}{cc}
                     \boldsymbol{V}_{11} & \boldsymbol{V}_{12} \\
                     \boldsymbol{V}_{21} & \boldsymbol{V}_{22} \\
                   \end{array}
                 \bigg),
\nonumber
\end{split}
\end{equation}
where $\boldsymbol{V}_{11} = v_{s,s}$, $\boldsymbol{V}_{12} = (v_{s,1},\ldots, v_{s,k})$, and $\boldsymbol{V}_{22}$ has elements $v_{i,j}$ for $i,j \in \{1, \ldots, k\}$.

According to West and Harrison (1997, p. 637), $(t_{s}| t_1 + t_{s},\ldots,t_k + t_{s})$ is normally distributed with mean and variance:
\begin{equation}
\begin{split}
E[t_{s}| t_1 + t_{s},\ldots,t_k + t_{s}]  &= \boldsymbol{m}_1 + \boldsymbol{V}_{12}\boldsymbol{V}_{22}^{-1}((t_1 + t_{s},\ldots,t_k + t_{s})' - \boldsymbol{m}_2) \\
\mathit{Var}[t_{s}| t_1 + t_{s},\ldots,t_k + t_{s}]  &= \boldsymbol{V}_{11} - \boldsymbol{V}_{12}\boldsymbol{V}_{22}^{-1}\boldsymbol{V}_{21}.
\nonumber
\end{split}
\end{equation}
Hence, $(t_{s}| p_1,\ldots,p_k)$ is normally distributed with mean and variance
\begin{equation}
\begin{split}
E[t_{s}|p_1,\ldots,p_k]  &= \boldsymbol{m}_1 + \boldsymbol{V}_{12}\boldsymbol{V}_{22}^{-1}\big((\sqrt{v_{n_1+n_{s}}}\Phi^{-1}(p_1)-\tau_1,\ldots,\sqrt{v_{n_k+n_{s}}}\Phi^{-1}(p_k)-\tau_1)' - \boldsymbol{m}_2\big) \\
\mathit{Var}[t_{s}| p_1,\ldots,p_k]  &= \boldsymbol{V}_{11} - \boldsymbol{V}_{12}\boldsymbol{V}_{22}^{-1}\boldsymbol{V}_{21}.
\nonumber
\end{split}
\end{equation}
Note that $\tau_1 = \sigma^2 \theta_0 / \sigma_0^2$ and $p_0 = F_0(\tau_1)=\Phi(\tau_1/\sqrt{v_0})$, which implies that $\tau_1 = \sqrt{v_0}\Phi^{-1}(p_0)$ and $\theta_0 = (\sigma_0^2/\sigma^2)\sqrt{v_0}\Phi^{-1}(p_0)$.

The second step in the proof is to apply Proposition~\ref{Proposition: Conjugate-Pair Ensemble with Private and Shared Information}. Let $L = -(k-1)\sqrt{v_{0}}\Phi^{-1}(p_0) + \sum_{i =1}^k \sqrt{v_{n_i+n_{s}}}\Phi^{-1}(p_i)$. According to Proposition~\ref{Proposition: Conjugate-Pair Ensemble with Private and Shared Information} and the derivation of Example~\ref{Example: Normal/Normal Pair} where $F_n(t) = \Phi(t/\sqrt{v_n})$, we have
\begin{equation}
\begin{split}
P(y=1|p_1,\ldots,p_k) &= \int_{-\infty}^{\infty} P(y=1|p_1,\ldots,p_k,t_{s}) f(t_{s}|p_1,\ldots,p_k) \ dt_{s} \\
&= \int_{-\infty}^{\infty} \Phi\bigg(\frac{L -(k-1)t_{s}}{\sqrt{v_{N_{k}+n_s}}}\bigg) f(t_{s}|p_1,\ldots,p_k) \ dt_{s} \\
&= \int_{-\infty}^{\infty} \bigg(1-\Phi\bigg(\frac{t_{s} - L/(k-1)}{\sqrt{v_{N_{k}+n_s}}/(k-1)}\bigg)\bigg) f(t_{s}|p_1,\ldots,p_k) \ dt_{s} \\
&= 1 - \int_{-\infty}^{\infty} \bigg(\int_{-\infty}^{t_{s}} \frac{k-1}{\sqrt{v_{N_{k}+n_s}}} \phi\bigg(\frac{z - L/(k-1)}{\sqrt{v_{N_{k}+n_s}}/(k-1)}\bigg) \ dz\bigg) f(t_{s}|p_1,\ldots,p_k) \ dt_{s} \\
&= 1 - \int \int_{z \leq t_{s}}  \frac{k-1}{\sqrt{v_{N_{k}+n_s}}} \phi\bigg(\frac{z - L/(k-1)}{\sqrt{v_{N_{k}+n_s}}/(k-1)}\bigg)  f(t_{s}|p_1,\ldots,p_k) \ dz dt_{s} \\
&= 1 - P(z \leq t_{s}|p_1,\ldots,p_k) = P(t_{s} \leq z|p_1,\ldots,p_k) =  P(t_{s} - z \leq 0|p_1,\ldots,p_k),
\nonumber
\end{split}
\end{equation}
where $(z|p_1,\ldots,p_k) \sim N(L/(k-1), v_{N_{k}+n_s}/(k-1)^2)$ and $z$ and $t_{s}$ are independent given $(p_1,\ldots,p_k)$. Consequently, $t_{s} - z \sim N( E[t_{s}|p_1,\ldots,p_k] - L/(k-1), \mathit{Var}[t_{s}| p_1,\ldots,p_k] + v_{N_{k}+n_s}/(k-1)^2)$, which gives us the result.

\subsection*{Lemma~1 and its Proof.}
The result below provides key properties of the linear combination of information states. We use these properties in the proof of Proposition~\ref{Proposition: Generalized Probit Ensemble} below.

\begin{lemma}[Linear Combination of Information States]
\begin{it}
Assume $\boldsymbol{x} \sim N(\boldsymbol{\mu}, \boldsymbol{\Sigma})$.  Then the following statements hold.
\begin{itemize}
\item[(i)] The distribution of $\alpha_0 + \sum_{i=1}^{k} \alpha_i x_i$ is normal with mean $m_0 = \alpha_0 + \sum_{i=1}^{k} \alpha_i \mu_i$ and variance $v_0 = \boldsymbol{\alpha}\boldsymbol{\Sigma}\boldsymbol{\alpha}' = \sum_{i =1}^k \sum_{j =1}^k \alpha_i \alpha_j \sigma_{ij}$.

\item[(ii)] The joint distribution of $\alpha_0 + \sum_{j \neq i}\alpha_j x_j$ and $\alpha_i x_i$ is normal with mean $\boldsymbol{a} + \boldsymbol{A}_i\boldsymbol{\mu}$ and variance $\boldsymbol{A}_i\boldsymbol{\Sigma}\boldsymbol{A}_i'$, where
\begin{equation}
\begin{split}
\boldsymbol{a} = \bigg(
                   \begin{array}{c}
                     \alpha_0 \\
                     0       \\
                   \end{array}
                 \bigg)  \quad \mathrm{and} \quad  \boldsymbol{A}_i = \bigg(
                   \begin{array}{ccccccc}
                     \alpha_1 & \cdots & \alpha_{i-1} & 0      & \alpha_{i-1} & \cdots & \alpha_k \\
                     0       & \cdots & 0           & \alpha_{i} & 0           & \cdots & 0 \\
                   \end{array}
                 \bigg).
\label{a vector and A matrix}
\nonumber
\end{split}
\end{equation}

\item[(iii)] The conditional distribution of $\alpha_0 + \sum_{j \neq i}\alpha_j x_j$ given $\alpha_i x_i$ is normal with mean $m_i = \theta_{-i} + b_{-i,i}b_{i,i}^{-1}(\alpha_i x_i - \theta_{i})$ and variance $v_i = b_{-i,-i} - b_{-i,i}b_{i,i}^{-1}b_{i,-i}$, where
\begin{equation}
\begin{split}
\bigg(\begin{array}{c}
                     \theta_{-i} \\
                     \theta_i    \\
                   \end{array}
                 \bigg) = \boldsymbol{a} + \boldsymbol{A}_i\boldsymbol{\mu}, \quad   \boldsymbol{B}_i= \bigg(
                   \begin{array}{ccccccc}
                     b_{-i,-i} & b_{-i,i}\\
                     b_{i,-i} & b_{i,i} \\
                   \end{array}
                 \bigg) = \boldsymbol{A}_i\boldsymbol{\Sigma}\boldsymbol{A}_i',
\label{b vector and B matrix}
\nonumber
\end{split}
\end{equation}
\begin{equation}
\begin{split}
b_{-i,-i} = \sum_{j \neq i} \sum_{j' \neq i} \alpha_j \alpha_{j'} \sigma_{jj'}, \quad b_{-i,i} = \sum_{j \neq i} \alpha_j \alpha_i \sigma_{ij}, \quad \mathrm{and} \quad \quad b_{i,i} = \alpha_i^2 \sigma_{ii}.
\nonumber
\end{split}
\end{equation}
\end{itemize}
A second expression for the variance $v_i$ is $v_i  = \sum_{j \neq i} \sum_{j' \neq i} \alpha_j \alpha_{j'} \sigma_{jj'} - \sigma_{ii}^{-1}\big(\sum_{j \neq i} \alpha_j \sigma_{ij}\big)^2$.
\end{it}
\end{lemma}

\noindent {\it Proof.}  Statements (i) and (ii) follow directly from results in West and Harrison (1997, p. 637)\nocite{west:harr:1997} on linear transformations of jointly distributed normal random variables. Statement (iii) follows from Statement (ii) and West and Harrison's (1997, p. 637)\nocite{west:harr:1997} result on conditional distributions of jointly distributed normal random variables.  The variance of $\alpha_0 + \sum_{i=1}^{k} \alpha_i x_i$ follows from the formula for the covariance of two linear transformations:
\begin{equation}
\begin{split}
v_0 = \mathrm{Cov}\bigg[\alpha_0 + \sum_{i=1}^{k} \alpha_i x_i,\alpha_0 + \sum_{i=1}^{k} \alpha_i x_i\bigg] = \sum_{i =1}^k \sum_{j =1}^k \alpha_i \alpha_j \mathrm{Cov}[x_i,x_j] = \sum_{i =1}^k \sum_{j =1}^k \alpha_i \alpha_j \sigma_{ij}.
\nonumber
\end{split}
\end{equation}
The elements of $\boldsymbol{B}_i$ follow from the same formula:
\begin{equation}
\begin{split}
b_{-i,-i} = \mathrm{Cov}\bigg[\alpha_0 + \sum_{j \neq i}\alpha_j x_j,\alpha_0 + \sum_{j \neq i}\alpha_j x_j\bigg] =  \sum_{j \neq i} \sum_{j' \neq i} \alpha_j \alpha_{j'} \mathrm{Cov}[x_j,x_{j'}] = \sum_{j \neq i} \sum_{j' \neq i} \alpha_j \alpha_{j'} \sigma_{jj'},
\nonumber
\end{split}
\end{equation}
\begin{equation}
\begin{split}
b_{-i,i} = \mathrm{Cov}\bigg[\alpha_0 + \sum_{j \neq i}\alpha_j x_j,\alpha_i x_i\bigg] =  \sum_{j \neq i} \alpha_j \alpha_i \mathrm{Cov}[x_j,x_i] = \sum_{j \neq i} \alpha_j \alpha_i \sigma_{ij},
\nonumber
\end{split}
\end{equation}
and $b_{i,i} = \mathrm{Cov}[\alpha_i x_i,\alpha_i x_i] = \alpha_i^2 \sigma_{ii}$, so that
\begin{equation}
\begin{split}
v_i  & =  b_{-i,-i} - b_{-i,i} b_{i,i}^{-1}b_{i,-i} =b_{-i,-i} - \frac{b_{-i,i}^2}{b_{i,i}} = \sum_{j \neq i} \sum_{j' \neq i} \alpha_j \alpha_{j'} \sigma_{jj'} - \frac{\big(\sum_{j \neq i} \alpha_j \alpha_i \sigma_{ij}\big)^2}{\alpha_i^2 \sigma_{ii}}  \\
     & =  \sum_{j \neq i} \sum_{j' \neq i} \alpha_j \alpha_{j'} \sigma_{jj'} - \frac{\big(\sum_{j \neq i} \alpha_j \sigma_{ij}\big)^2}{\sigma_{ii}}.
\nonumber
\end{split}
\end{equation}

\subsection*{Proof of Proposition~\ref{Proposition: Generalized Probit Ensemble}.}  The conditional probability of $(y|\alpha_i x_i)$ is given by
\begin{equation}
\begin{split}
P(y=1|\alpha_i x_i) = \int_{-\infty}^{\infty} f(x|\alpha_i x_i) F_{z_0}(x + \alpha_i x_i) \, dx,
\nonumber
\end{split}
\end{equation}
where $x = \alpha_0 + \sum_{j \neq i} \alpha_j x_j$, $f(x|\alpha_i x_i)$ is the normal density according to the joint normality assumption and Lemma~1(iii), and $P(y=1|\boldsymbol{x}') = F_{z_0}(x + \alpha_i x_i)$ comes from the generalized linear model assumption. We can rewrite this expression as
\begin{equation}
\begin{split}
P(y=1|\alpha_i x_i) &= \int_{-\infty}^{\infty} \frac{1}{\sqrt{v_i}} \phi\bigg(\frac{x-m_i}{\sqrt{v_i}} \bigg) \bigg(\int_{-\infty}^x f_{z_0}(z + \alpha_i x_i)\,dz \bigg) \, dx \\
                   &= \int \int_{z \leq x} \frac{1}{\sqrt{v_i}} \phi\bigg(\frac{x-m_i}{\sqrt{v_i}} \bigg) f_{z_0}(z + \alpha_i x_i)\,dz dx = P(z \leq x) = P(z - x \leq 0),
\label{Equation: Evaluating the calibration condition}
\end{split}
\end{equation}
where on the right-hand side of the third equality, $z$ and $x\sim N(m_i,v_i)$ become random variables.  By the assumption about the link function, the random variable $z = \psi z_0 + l_i$ is from a location-scale family with the standard cdf $F_{z_0}$, location parameter $l_i = -\alpha_i x_i$, and scale parameter $\psi =1$. Since the last integrand involves a product of these random variable's densities, $z$ and $x$ are independent, as are $z$ and $-x \sim N(-m_i,v_i)$. The random variable $-x$, being normally distributed, is also from a location-scale family where $-x = \sqrt{v_i} x_0 - m_i$ and $x_0 \sim N(0,1)$.  We can express the last probability in \eqref{Equation: Evaluating the calibration condition} as
\begin{equation}
\begin{split}
P(z - x \leq 0) = P(z_0 + l_i +\sqrt{v_i} x_0 - m_i \leq 0)  = P(z_0 +\sqrt{v_i} x_0 \leq m_i - l_i).
\nonumber
\end{split}
\end{equation}
Let $F_{z_0 +\sqrt{v_i} x_0}(u) = P(z_0 +\sqrt{v_i} x_0 \leq u)$, the cdf of the sum of the independent random variables $z_0$ and $\sqrt{v_i} x_0$.  Because $m_i-l_i= \theta_{-i} - b_{-i,i}b_{i,i}^{-1}\theta_{i} + (b_{-i,i}b_{i,i}^{-1}+1)\alpha_i x_i$, we have
\begin{equation}
\begin{split}
P(y=1|\alpha_i x_i) = F_{z_0 +\sqrt{v_i} x_0}(\theta_{-i} - b_{-i,i}b_{i,i}^{-1}\theta_{i} + (b_{-i,i}b_{i,i}^{-1}+1)\alpha_i x_i),
\nonumber
\end{split}
\end{equation}
which is strictly monotonic in $x_i$.

By Bayes' Theorem and a change of variable, we have that $P(y=1|q) = P(y=1|x_i)$ where $q = \alpha_i x_i$:
\begin{equation}
\begin{split}
P(y=1|q) &= \frac{P(y=1)f_{q|y=1}(q|y=1)}{f_{q}(q)} = \frac{P(y=1)f_{q|y=1}(\alpha_i x_i|y=1)\alpha_i}{f_{q}(\alpha_i x_i)\alpha_i} \\
           &= \frac{P(y=1)f_{x_i|y=1}(x_i|y=1)}{f_{x_i}(x_i)} = P(y=1|x_i).
           \nonumber
\label{Equation: Calibration proof}
\end{split}
\end{equation}
Then by the assumption that $p_i = P(y=1|x_i)$, we have that $p_i = P(y=1|\alpha_i x_i)$. Consequently,
\begin{equation}
\begin{split}
p_i &= F_{z_0 +\sqrt{v_i} x_0}(\theta_{-i} - b_{-i,i}b_{i,i}^{-1}\theta_{i} + (b_{-i,i}b_{i,i}^{-1}+1)\alpha_ix_i) \\
\iff \quad \alpha_ix_i
&= \frac{F_{z_0 +\sqrt{v_i} x_0}^{-1}(p_i)+b_{-i,i}b_{i,i}^{-1}\theta_{i}-\theta_{-i}}{(b_{-i,i}b_{i,i}^{-1}+1)}.
\nonumber
\end{split}
\end{equation}
Substitution of this last expression into the assumed generalized linear model gives us
\begin{equation}
\begin{split}
\alpha_0 + \sum_{i=1}^k \alpha_i x_i = \alpha_0 + \sum_{i=1}^{k}\frac{b_{-i,i}b_{i,i}^{-1}\theta_{i}-\theta_{-i}}{(b_{-i,i}b_{i,i}^{-1}+1)} + \sum_{i=1}^{k}\beta_iF_{z_0 +\sqrt{v_i} x_0}^{-1}(p_i),
\nonumber
\end{split}
\end{equation}
where
\begin{equation}
\begin{split}
\beta_i = \frac{1}{b_{-i,i}b_{i,i}^{-1}+1} = \frac{b_{i,i}}{b_{-i,i} + b_{i,i}} = \frac{\alpha_i^2 \sigma_{ii}}{\sum_{j=1}^k \alpha_j \alpha_i \sigma_{ij}}  = \frac{\alpha_i \sqrt{\sigma_{ii}}}{\sum_{j=1}^k \alpha_j \sqrt{\sigma_{jj}}\rho_{ij} }.
\nonumber
\end{split}
\end{equation}
Because $\theta_i = \alpha_i \mu_i$ and $\theta_{-i} = \alpha_0 + \sum_{j\neq i}\alpha_j \mu_j$, which implies that $\theta_i + \theta_{-i} = m_0$ from Lemma~1, we have that
\begin{equation}
\begin{split}
\alpha_0 +\sum_{i=1}^{k}\frac{b_{-i,i}b_{i,i}^{-1}\theta_{i}-\theta_{-i}}{(b_{-i,i}b_{i,i}^{-1}+1)} &= \alpha_0 +\sum_{i=1}^{k} (1-\beta_i)\theta_{i} - \sum_{i=1}^{k}\beta_i \theta_{-i}  = \alpha_0 +\sum_{i=1}^{k}\theta_{i} - \sum_{i=1}^{k}\beta_i (\theta_i + \theta_{-i}) \\
&= \alpha_0 +\sum_{i=1}^{k}\alpha_i\mu_{i} - \sum_{i=1}^{k}\beta_i m_0 = \bigg(1 - \sum_{i=1}^{k}\beta_i\bigg) m_0.
\nonumber
\label{Intercept}
\end{split}
\end{equation}
From the expressions above, we can rewrite expert $i$'s reported probability as follows:
\begin{equation}
\begin{split}
p_i &= F_{z_0 +\sqrt{v_i} x_0}(\theta_{-i} - b_{-i,i}b_{i,i}^{-1}\theta_{i} + (b_{-i,i}b_{i,i}^{-1}+1)\alpha_ix_i) \\
    &= F_{z_0 +\sqrt{v_i} x_0}(\theta_{i} + \theta_{-i} - (b_{-i,i}b_{i,i}^{-1} + 1)\theta_{i} + (b_{-i,i}b_{i,i}^{-1}+1)\alpha_ix_i) \\
    &= F_{z_0 +\sqrt{v_i} x_0}(\theta_{i} + \theta_{-i} + (b_{-i,i}b_{i,i}^{-1}+1)\alpha_i (x_i -\mu_i)) \\
    &= F_{z_0 +\sqrt{v_i} x_0}\bigg(\alpha_0 + \sum_{i=1}^{k} \alpha_i \mu_i + \beta_i^{-1}\alpha_i (x_i -\mu_i)\bigg).
\nonumber
\end{split}
\end{equation}

Finally, $P(y=1) = \int_{-\infty}^{\infty} f(w) F_{z_0}(w) \, dw$ where $w = \alpha_0 + \boldsymbol{\alpha}\boldsymbol{x}$ and $f(w)$ is the normal density according to the joint normality assumption, the generalized linear model assumption, and Lemma~1(i). This expression can be simplified:
\begin{equation}
\begin{split}
P(y=1) &= \int_{-\infty}^{\infty} \frac{1}{\sqrt{v_0}} \phi\bigg(\frac{w-m_0}{\sqrt{v_0}} \bigg) \bigg(\int_{-\infty}^w f_{z_0}(z)\,dz \bigg) \, dw = \int \int_{z \leq w} \frac{1}{\sqrt{v_0}}\phi\bigg(\frac{w-m_0}{\sqrt{v_0}} \bigg) f_{z_0}(z)\,dz dw \\
                   &= P(z_0 \leq w) = P(z_0 - w \leq 0),
\nonumber
\end{split}
\end{equation}
where on the right-hand side of the third equality, $z_0$ and $w$ become independent random variables. Because $-w = \sqrt{v_0} x_0 - m_0$ and $x_0 \sim N(0,1)$, we can express the last probability as $P(z_0 - w \leq 0) = P(z_0 +\sqrt{v_0} x_0 \leq m_0)$. Let $F_{z_0 +\sqrt{v_0} x_0}(u) = P(z_0 +\sqrt{v_0} x_0 \leq u)$. Then $m_0 = F_{z_0 +\sqrt{v_0} x_0}^{-1}(p_0)$ and $\beta_0 = 1 - \sum_{i=1}^{k}\beta_i$.

\bibliographystyle{ormsv080}
\bibliography{bayesian-ensembles-of-binary-event-forecasts}

\end{document}